\documentclass[aps,twocolumn,floats,showpacs,superscriptaddress]{revtex4}
\pdfoutput=1
\usepackage{graphics,dcolumn,float}
\usepackage{graphicx}
\usepackage{dcolumn}
\usepackage{bm}
\usepackage{amsmath}
\usepackage{amssymb}
\usepackage{color}
\usepackage{soul}
\usepackage{epstopdf}
\begin{document}
\preprint{APS/123-QED}
\title{Spin to pseudo-spin conversion in graphene-like systems: A Kubo formalism including vertex corrections}

\author{R. Baghran}
\affiliation{Department of Physics, Shahid Beheshti University, 1983963113 G.C., Evin
	Tehran, Iran}
\author{M.M. Tehranchi}
\affiliation{Department of Physics, Shahid Beheshti University, 1983963113 G.C., Evin
	Tehran, Iran}
\author{A. Phirouznia}%
\email[Corresponding author's Email: ]{phirouznia@azaruniv.ac.ir}
\affiliation{Department of Physics, Azarbaijan Shahid Madani
University, 53714-161, Tabriz, Iran}
\affiliation{Condensed Matter
Computational Research Lab. Azarbaijan Shahid Madani University,
53714-161, Tabriz, Iran}

\date{\today}
\begin{abstract}
Spin to pseudo-spin conversion by which spin population imbalance converts to non-equilibrium pseudo-spin density in Dirac systems has been investigated particularly for graphene and insulator phase of silicene.
Calculations have been performed within the Kubo approach and by taking into account the vertex correction. Results indicate that spin converts to pseudo-spin in either graphene or silicene that identified to come from the spin-orbit interactions. The response function of spin to pseudo-spin conversion is weakened several orders of magnitude by vertex correction of impurities in graphene, however, this conversion is  strengthened in insulator silicene. In addition, in the case of silicene, results are indicative of an obvious change in the mentioned response function as a result of the change in band-topology which can be observed by manipulation of external electric field, vertically applied to the system surface. At the critical electric field in which the topological phase transition for silicene nano-ribbon has been observed, response function  changes abruptly. Interconversion between the quantum numbers could provide a field for information and data processing technologies.
\end{abstract}


\maketitle

\section{INTRODUCTION}
Benefits like low-power operation and simple qubit-based spin language for applications in processing and data storage, has made spintronics devices more attractive than electronics ones in recent years \cite{1,2}. On the other hand, one of the most substantial scientific challenges in spintronics is manipulation of spin configurations by charge transport. In fact, one of the central goals in spintronics is introducing an efficient mechanism for electrically controlling both  generation and detection of spin current and spin accumulation   \cite{3,4}.

Meanwhile, the conversion between different quantum numbers with different dephasing and diffusive lengths could be employed in the future data transfer technologies. Coupling which could be established between the spin and other quantum numbers can effectively control the spin relaxation time \cite{51}. 
In this way, capability of charge-spin interconversion which arises from intrinsic spin-orbit interaction (SOI) has become one of the key phenomena for spintronics recently. These phenomena have been realized by two mechanisms, spin Hall effect (SHE)  \cite{5,6,7,8,9,10} and Edelstein effect (EE) some times known as inverse spin galvanic effect (ISGE) \cite{11,12,13,14,15,16}.
The SHE and its Onsager reciprocal inverse, (inverse spin Hall effect) come from an interconversion between charge current and transverse electronic spin current\cite{17,18}.  These effects have widely been
studied in heavy metal layers \cite{19,20,21}, semiconductors and two-dimensional materials \cite{22}, spin-valve structures \cite{23} and superconductors \cite{24}. 

The SHE has been studied in disordered materials by taking into account vertex corrections within a diagrammatic based framework \cite{25,26} of perturbation theory. The SHE and inverse SHE (ISHE), could be considered as different methods for generation and detection of pure spin current that carries a net angular momentum \cite{4,8}.
The EE (ISGE) and its inverse (IEE) (or equivalently spin galvanic effect (SGE)) were first propounded by Ivchenko and Pikus \cite{27}, observed in $Te$ \cite{28} and studied theoretically in the two dimensional electron gas (2DEG) with broken inversion symmetry, in the presence of dresselhaus and Rashba spin-orbit interactions (SOIs) \cite{29,30,31}. 

SOIs lead to interconversion between the charge current   and  non-equilibrium spin density  or spin population imbalance due to spin-momentum locking which arises from the lack of inversion symmetry that results in Rashba SOI, in low- dimensional systems such as semiconductors, and spin-momentum coupling in surface of topological insulators (TI)s \cite{32,33,34} and Weyl semi-metals \cite{36}, oxide interfaces \cite{16,36} and two dimensional systems \cite{37,38,39,40,41,42}. 

Spin-momentum coupling has also  been realized at effective Hamiltonian of TIs. Akzyanov \cite{52} studied the spin conductivity of the surface states in a thin film of a TIs within Kubo formalism where it has been shown that, these structures are promising materials for spintronic applications. 

It could be highlighted that in Dirac systems electron's Bloch wave function could be propagated on two in-equivalent sublattices. Hence
electrons possess two component quantum degree of freedom apart from their orbital ones called pseudo-spin, which could be related to a real angular momentum and physically measurable effects  \cite{43,44}.  The pseudo-spin concept is very substantial in Dirac systems because many physical processes in these systems are completely pseudo-spin-dependent and could be  understood using this framework \cite{45}.  

Pseudospin signatures have been detected in several experiments \cite{46,47,48}. As the Dirac equation in graphene-like materials provides pseudo-spin-momentum coupling. It could be inferred that the pseudo-spin-momentum locking opens a very tempting way to realization of electric-based  manipulation of pseudo-spin polarization for information and data processing applications. As Pesin et al. predicted in 2012, that charge current in single-layer graphene is accompanied with pseudo-spin currents \cite{49}. In 2014, Chen et al. \cite{50} investigated the possibility of extracting pseudo-spin polarization by means of electric field assisted electron emission. Spin manipulation has been studied in graphene by chemically induced pseudo-spin polarization \cite{51}. In this case certain type of impurities such as fluorine adatoms, locally break sublattice symmetry. After all, pseudo-spin manipulation was rarely investigated for graphene in recent years.
\\
Spin-charge interconversion which deals mainly with researches have been made in the field of SHE and EE covers a wide range of works from metal/oxide to Weyl semimetals and quantum wells \cite{35,52,53,54,55,56,57,58,59,60,61,62}. Strong enhancement of the Edelstein effect in $f$-electron systems close to the coherence temperature has also been reported  which could be explained by a coupling
between the conduction electrons and the localized $f-$electron \cite{58}.
\\
 Unlike the spin and charge current correlations which has been widely investigated, it seems that little attention has been paid to realization of pseudo-spin polarization or its manipulation by other quantum numbers such as spin. Therefore, pseudo-spin correlation with other observables such as spin pseudo-spin interconversion deserves more investigations.  
\\
Current study presents a mechanism for spin pseudo-spin interconversion in realistic disordered graphene-like systems within the Dirac point approximation. In the context of Kubo formalism and by taking in to account the vertex corrections it can be shown that the non-equilibrium spin density converts to pseudo-spin population imbalance (PPI) i.e. pseudo-spin polarization. Spin to pseudo-spin conversion could be employed in detection and measurement of pseudo-spin current/density. Since spin polarization can easily be measured with devices that are available already. This means that the information which transfers with one of these quantum numbers can be translated into another in data processing devices.
\section{Theory and approach }
The graphene-like Dirac-fermion structures with two $A$ and $B$ sublattices is generally described by the  Hamiltonian in the basis of $\{A\uparrow, A\downarrow, B\uparrow, B\downarrow \}$ as \cite{63}
\begin{eqnarray}
\label{Hamiltonian}
H^{\eta}_D &=& \hbar  v_F\left(\eta  k_x \tau _x+k_y
\tau _y\right)\nonumber \\&&+ \eta \tau _z \left(a \lambda _{ R2}\left(k_y \sigma _x-k_x \sigma
_y\right)\right)+\eta\lambda _{so}  \tau _z\sigma _z \nonumber\\&&-l E_z \tau _z+\frac{1}{2} \lambda _{R1} \left(\eta  \tau _x \sigma
_y-\sigma _x \tau _y\right).
\end{eqnarray}
In which, $\hbar v_F=\frac{\sqrt{3}}{2}at$  where $a$ is lattice constant and $t$ is the first nearest neighbors hopping energy, $\tau= (\tau_x, \tau_y, \tau_z)$ refers to sublattice degree of freedom  called pseudo-spin which could be represented by well-known Pauli matrices similar to spin operator $\sigma  =(\sigma_x,\sigma_x,\sigma_x)$. Where, $\sigma_\alpha$ being the $\alpha$-component of the electron spin.
\\

In-plane components of pseudo-spin operator  $\tau_{x},~\tau_{y}$  represent nearest neighbors electron hopping between two in-equivalent sublattices $A$ and $B$, meanwhile, the out of plane operator $\tau_z$ is indicates sublattice quantum number $A$ or $B$. Each of band electrons have either of $A$ or $B$ sublattice Bloch wave function denoted by $\psi_{A}$ and $\psi_{B}$. $\psi_{A/B}$ remains invariant under the action of $\tau_z$  meanwhile, in plane pseudo-spin operators $\tau_{x(y)}$ changes the state of electrons wave function from $\psi_A$  to $\psi_B$  and vice versa. $\eta$ is the valley index  that gets $+1$, $-1$ values for the $K$ and $K'$ points respectively, $\lambda_{so}$  indicates the strength of intrinsic SOI, $l$ is the buckling hight which can be considered as a parameter that measures structural inversion asymmetry (SIA). This asymmetry results in intrinsic Rashba SOI with a strength given by $\lambda_{R2}$
in buckled graphene-like systems such as silicene while identically vanishes in the case  of flat graphene sheet. Silicene has more stronger intrinsic SOI than graphene in which the intrinsic SOI is very weak  due to the small radius of carbon atoms. The strength of externally induced inversion asymmetry that leads to extrinsic Rashba SOI is identified by  $\lambda_{R1}$ and finally, the electric field $E_z$ is normally applied to the plane of graphene-like systems. 
\\

In the case of silicene, there is a relatively notable band gap of $1.55$ meV obtained by first principle studies \cite{63} in comparison with the gap-less graphene, and applied electric field $E_z$  can be employed to band-gap tuning in silicene. The band gap energy, $\Delta$, is given by 
\begin{eqnarray}
\Delta = 2|-\eta \lambda_{so}s_z+lE_z|,
\end{eqnarray}
where $s_z=\pm 1$ is normal component of electron spin. Based on series of works made on buckled silicene it has been shown that  at a critical normal field, $E_c=\pm 17(meV/\AA)$ the gap closes \cite{63,64}.  This can be interpreted as a point of band-topology change for silicene by which silicene’s gap opens at $K$ point for up spins and closes in $K'$ for down spins. This can be interpreted as spin-valley locking  in the presence of normal electric field  in silicene. This means that each of valleys has also its own normal spin polarization when $E_z\neq 0$ . For finite width silicene nano-ribbon the change of band topology by external field results in  topological phase transition by which, for $|E_z|<E_c$ silicene nano-ribbon becomes a topological insulator \cite{63}.

The Edelstein effect indicates that charge current can be converted to spin population imbalance. The extrinsic/intrinsic Rashba SOI terms in the Hamiltonian (Eq. 1) which arises from lack of inversion symmetry and makes spin-momentum locking, actually guarantees non-vanishing Edelstein response function in graphene and graphene-like structures. In the case of graphene which has a completely planar structure with zero buckling, $l=0$, there is no intrinsic Rashba SOI ($\lambda_{R2}=0$) meanwhile due to the extrinsic Rashba SOI which arises from external gate or substrate given by $H_{R1}=\frac{\lambda_{R1}}{2}(\eta\tau_x\sigma_y-\tau_y\sigma_x)$  a shift of distribution function  in the $k_x$-direction  $\delta k_x$, which arises as a result of an electric field along the $x$-axis, leads to non-vanishing current $J_x$ or $<\tau_x>\neq 0$. This means that the shift of distribution function by  $\delta k_x$, leads to change of the effective magnetic field of $y$ component, $\delta B^{eff}_{my}$. Accordingly, a non-equilibrium spin polarization, $\delta \sigma^y$ can be achieved. Therefore, the EE can be obtained by extrinsic Rashba SOI \cite{16}.

In the other words the EE in graphene could be explained by definition of an effective magnetic field as                                     
\begin{eqnarray}
\mathbf{B}^{eff}_m = (-\frac{\lambda_{R1}}{2}\tau_y,\frac{\lambda_{R1}}{2}\eta\tau_x,\eta\lambda_{so}\tau_z).
\end{eqnarray}
A shift in one of the in-plane component of $B^{eff}_m$ consequently  leads to a net spin polarization at the same direction. Applying  an electric field of $E_{x(y)}$ leads to $\delta k_{x(y)}$ shift in distribution function. This shift ($\delta k_{x(y)}$) that results in  $<\tau_{x(y)}>\neq 0$                 and eventually  $<\tau_{x(y)}>\neq 0$  gives rise to $\delta B^{y(x)}$  which makes a net spin polarization in $y(x)$ direction.

Edelstein conductivity could be introduced as $E_{x(y)}=\sigma_{EE}S^{y(x)}$ and its Onsager reciprocal $S^{y(x)}=\sigma_{IEE}E_{x(y)}$ in which $\sigma_{IEE}$ indicates the conductivity of inverse Edelstein effect (IEE) that refers to conversion of non-equilibrium spin population to the charge current \cite{16}. 

In the case of other buckled graphene-like structures and in the presence of intrinsic Rashba SOI, which is given by $H_{R2}=a\lambda_{R2}(k_y\sigma_x-k_x\sigma_y)$ a $k$-space shift $\delta k_x$ of distribution function  directly leads to $\delta \sigma^y$, so longitudinal charge current $J_x$ converts directly to transverse spin population imbalance $\delta \sigma^y$  and contributes in the EE. 

In the present study a new type of conversion which takes place between the other quantum numbers of Dirac fermions have been introduced and analyzed: \emph{the spin to pseudo-spin conversion}. In what follows, it can be realized that under such circumstances, how the pseudo-spin polarization could arise as a result of the pseudo-magnetic field of the low energy effective Hamiltonian.

One can unify all terms of the Hamiltonian $H_D$ in term of the effective pseudo-magnetic field as a Zeeman-like interaction therefore, Hamiltonian of the system gets replace with  
\begin{eqnarray}
H_{eff} = \mathbf{B}^{eff}_{pm}\cdot\bf{\tau},
\end{eqnarray}       
by definition effective pseudo-magnetic field is given as
\begin{eqnarray}
\mathbf{B}^{eff}_{pm} = (\hbar v_F \eta k_x+\eta\frac{\lambda_{R1}}{2}\sigma_y,\hbar v_F k_y-\frac{\lambda_{R1}}{2}\sigma_x,
\nonumber\\ \eta \lambda_{so}\sigma_z-l E_z +\eta a \lambda_{R2}(k_y\sigma_x-k_x\sigma_y))
\end{eqnarray}                        
The effective pseudo-magnetic field defined in (2.5) could be simplified for graphene as 
\begin{eqnarray}
\label{2.6}
\mathbf{B}^{(\eta)}_{Gpm} =(\hbar v_F \eta k_x+\eta\frac{\lambda_{R1}}{2}\sigma_y,\hbar v_F k_y-\frac{\lambda_{R1}}{2}\sigma_x, \eta \lambda_{so}\sigma_z).\nonumber\\
\end{eqnarray}
In the case of silicene, the effective pseudo-magnetic field could be written as
\begin{eqnarray}
\label{2.7}
\mathbf{B}^{(\eta)}_{Spm} = (\hbar v_F \eta k_x+\eta\frac{\lambda_{R1}}{2}\sigma_y,\hbar v_F k_y-\frac{\lambda_{R1}}{2}\sigma_x,
\nonumber\\ \eta \lambda_{so}\sigma_z-l E_z +\eta a \lambda_{R2}(k_y\sigma_x-k_x\sigma_y)).
\end{eqnarray}
An applied magnetic  field along the $z$-direction generates a spin polarization of $\delta\sigma^{\eta}_z=<\sigma_z>^{\eta}_B-<\sigma_z>^{\eta}_{B=0}$ by Zeeman interaction directly which leads to an effective pseudo magnetic field of $\delta{B}^{(\eta)}_{pm}=\eta \lambda_{so}\delta \sigma^{\eta}_z \hat{z}$  which eventually results in PPI identified by $\delta \tau_z\neq 0$. This means that the non-equilibrium spin accumulation results in valley dependent non-equilibrium pseudo-spin polarization at each valley. It seems that the net pseudo-spin polarization falls to zero since the effective pseudo-magnetic field has opposite sign at each valley. However, this is rather unsurprising given the fact that $k$-space average of  $\delta\sigma^{\eta}_z$ is not the same for each of the valleys. As depicted in Figs. 1 and 2 effective pseudo-magnetic field shows different form at $K$ and $K'$ valleys in graphene.  On the other hand, in the case of insulator silicene when $E_z\neq 0$, spin valley locking \cite{63} results in oppositely spin polarization of non-equivalent valleys. Then if we assume that the spin-valley locking results in spin up and spin down polarizations in $K$ and $K'$ valleys respectively, therefore a normal magnetic field that increases the spin up population at $K$ Dirac point decreases the down spin population at $K'$ point. Therefore, it can be realized that in the case of silicene $\delta\sigma^{K}_z=-\delta\sigma^{K'}_z$ and non-equilibrium induced pseudo magnetic field of different valleys cannot cancel out each other. Accordingly there is net non-vanishing pseudo-magnetic field  $\delta{B}_{pm}=\sum_{\eta}\delta{B}^{(\eta)}_{pm}\neq 0$. 
Remarkably, the intrinsic SOI is of key importance in spin to pseudo-spin response function, meanwhile, the physical parameters behind this process are not exactly the same for graphene in the metallic phase and other buckled graphene-like structures at insulator regime.
\\ 
\begin{figure}[h]
	\includegraphics[width=1.0\linewidth]{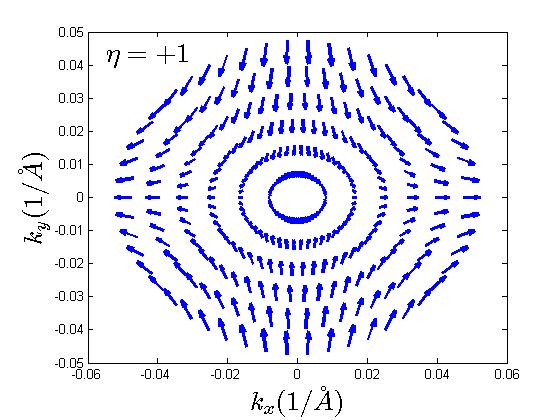}
	\centering
	\caption{(Color online) Pseudo-magnetic vector field of graphene around the $K$ point.  \label{vfield_K}}
\end{figure}
\begin{figure}[h]
	\includegraphics[width=1.0\linewidth]{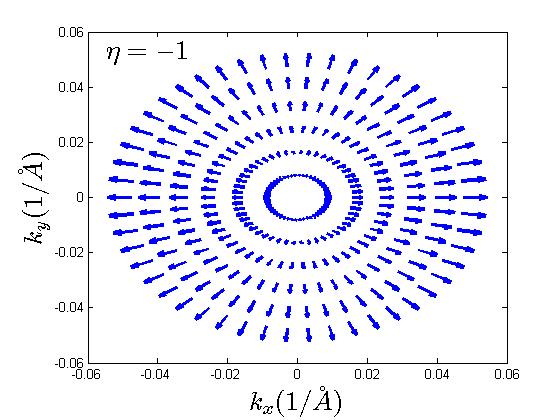}
	\centering
	\caption{(Color online) Pseudo-magnetic vector field of graphene around the $K'$ point.  \label{vfield_Kp}}
\end{figure}

Although, the simple picture of the effective spin induced pseudo magnetic field seems to provide a reasonable explanation of spin to pseudo-spin conversion via the  intrinsic SOI, however, it can be shown that for low energy Dirac fermions, nonzero response function of spin to pseudo-spin conversion crucially depends on the presence of Rashba coupling. As The numerical results indicate, presence of the Rashba interaction plays an important role in effectiveness of the intrinsic SOI in this process. At zero Rashba coupling strength and  close to the Dirac points, low energy Hamiltonian of Dirac fermions is reduced to $ H_D \simeq \hbar v_F\left(\eta  k_x \tau _x+k_y
\tau _y\right) +\eta \lambda_{so}\tau_z \sigma_z$ that commutes with  spin operator. In this case effective magnetic field is oriented vertically to plane of the system.\\ Spin-pseudo-spin response function of pure system is given by
\begin{eqnarray}
\chi^{(0)}_{\tau_z s_z} =\sum_{\alpha, \alpha'}\frac{<\alpha|\tau_z|\alpha'><\alpha'|s_z|\alpha>}{E_{\alpha}-E_{\alpha'}+\hbar\omega+ i \epsilon} (f(E_{\alpha'}) - f(E_{\alpha})),\nonumber\\
\end{eqnarray}
where $|\alpha>$ and $E_{\alpha}$ are the unperturbed eigen-states and eigen-values respectively and $f(E_{\alpha})$ is the Fermi distribution function and $\epsilon $ is a small positive. According to the above expression,
it can easily be realized that when the motive perturbation (here $s_z$) commutes with Hamiltonian of the system, response function ($\chi^{(0)}_{\tau_z s_z}$) identically vanishes. In the presence of the Rashba interaction effective magnetic field deviates from the normal direction since the Rashba interaction introduces new in-plane effective field for spin and even pseudo-spin moments. Therefore in this case motive force that associated with $\sigma_z$ operator does not commute with Hamiltonian which accordingly results in non-vanishing response function. This fact can easily be understood if we consider that the perturbations commuting with the Hamiltonian cannot disturb its eigenstates and contribute in the non-equilibrium processes. This would suggest that the Rashba interaction has fundamental impact on effectiveness of the intrinsic SOI in spin pseudo-spin response function.    
     
Among the graphene-like structures, buckled configurations where the sub-lattice symmetry has been broken are the natural candidates for pseudo-spin polarization generator materials. In this case normal electric field appears actually as a pseudo-magnetic field which comes out as a pseudo-Zeeman term, $lE_z\tau_z$, in the Hamiltonian of buckled graphene like systems. In fact, at equilibrium in the case of buckled structures normal electric field generates pseudo-spin polarization. Meanwhile, in the current study it has been shown that how non-equilibrium pseudo-spin polarization could be produced in Dirac materials by normal magnetic filed. 

The influence of impurities can be considered at different levels i.e. up to the Born approximation that modifies the bare retarded and advanced Green's function or by normalizing the response function via the vertex corrections. These two level changes have different contributions on response function.      
\\

The spin to pseudo-spin conversion can be expressed by bare Kubo response function of $\sigma_z$ to $\tau_z$ shown in Fig. 3   denoted by $\chi_{\sigma_z\tau_z}$ can be formulated as \cite{26,30,65}
\begin{eqnarray}
\chi^\eta_{\sigma_z\tau_z}= \int \frac{d^2k}{(2\pi)^2} Tr[\tau_z G^{R(\eta)}_{0}(E_F,k) \sigma_z G^{A(\eta)}_{0}(E_F,k)].\nonumber\\
\end{eqnarray}
Where $G^{A/R}_{0}$  are the advanced/retarded  un-dressed Green’s functions in the absence of disorders given by,
\begin{eqnarray}
G^{A/R(\eta)}_{0}(E,k)=[E\times I- \hat{H}^{\eta}_D \mp i0^{+} ]^{-1},
\end{eqnarray}
where $I$ is $4\times4$ identity matrix. 
In the presence of impurities with the typical potential $V_{im}$, a level broadening is introduced  by the imaginary part of the self-energy,  $\Sigma$.
\begin{figure*}[t]
	\includegraphics[ width=0.80\linewidth ]{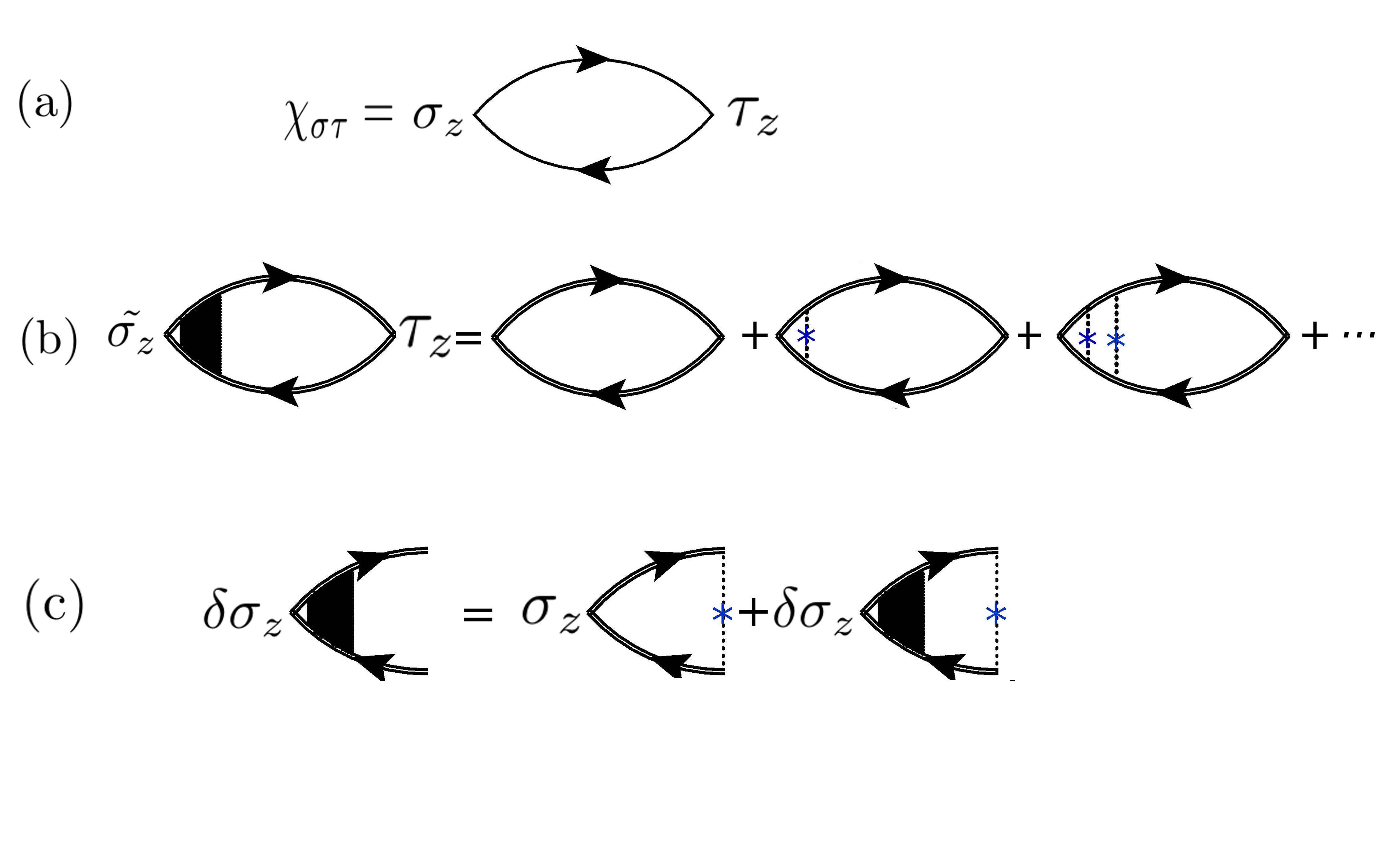}
	\caption{(Color online) Feynman bubble diagrams; the graphical representation of spin to pseudo-spin response function. Single lines are undressed Green's functions of the sample while double lines denote the dressed Green's functions. Cross shape blue points refer to impurity and dashed lines show scattering from impurity potential\cite{27}. a) Unperturbed response function of the spin and pseudo-spin.  b) Ladder diagrams correspond to the corrections of the response function which come from successive interactions of electron-hole pair with impurities that leads to normalized vertex spin operator. c) The diagrammatic approach of electron-hole pair interactions with impurities results in iterative equation of vertex correction.   \label{diagram}}
\end{figure*}

Dyson equation for Green’s function  in term of disorder averaged self- energies reads
\begin{eqnarray}
\Sigma^{A/R}&=& <V_{im}+V_{im}G^{A/R}_{0}V_{im}>_{dis} \\ 
\hat{H}&=&\hat{H}_D +\Sigma \\
G&=&G_{0}+G_{0}\Sigma G_{0}+G_{0}\Sigma G_{0}\Sigma+...\nonumber\\
&=& G_{0}+G_{0}\Sigma G. 
\end{eqnarray}

Where $<>_{dis}$ in (10) refers to disorder configurational average. Real part of the self-energy introduces a energy shift which could be ignored by redefinition of eigen energies and the imaginary part of the self-energies within the Born approximation reads $Im \Sigma^{A/R}=\mp 2i/\tau_{im}$. Here, $\tau_{im}$ is the relaxation time which directly relates to level broadening concept. Then the advanced/retarded dressed (effective) Green's functions could be written as 
\begin{eqnarray}
G^{A/R(\eta)}(E,k)=[E\times I-\hat{H}^{\eta}_D \mp \frac{2i}{\tau_{im}} ]^{-1},
\end{eqnarray}
$G^{A/R}$ indicates dressed (effective) Green’s function.
\\

The impurity potential $V_{im}$ is chosen to be a spin-orbit like scatterer as\cite{27}
\begin{eqnarray}
V_{im} = u_{0}R^{2} \displaystyle\sum_{j} M_{4\times 4} \delta (\vec{r}-\vec{r_{j}}).
\end{eqnarray}
In which summation runs over the position of impurities, $\mathbf{r}_j$, $u_0$ is the power of impurity potential, $R$ is length scale refers to impurity potential range and  $M_{4\times4}$ is the scattering matrix which has the functional form of intrinsic SOI and refers transitions made by impurities in $|\tau_z>\otimes|\sigma_z>$ space can be presented by
\begin{eqnarray}
\label{2.16}
M =  \tau_{z} \sigma_{z}.
\end{eqnarray}
Accordingly, the intrinsic SOI of both host atoms that periodically repeated over the structure and disorder atoms which randomly distributed among the periodic texture have the same functional form in pseudo-spin-spin space.

The scattering relaxation time within the first Born approximation is given as
\begin{eqnarray}
\label{2.17}
\frac{1}{\tau^{\lambda}_{im}(k)} &=& \frac{2\pi}{\hbar}\sum_{{\lambda'}{k'}} |<k\lambda|V_{im}|{k'} {\lambda'}>|^2 \times\\ && \delta (E_{k\lambda}-E_{k' \lambda'}) (f(E_{k\lambda}) - f(E_{k' \lambda'}) )\nonumber
\end{eqnarray}
Where $(\lambda,\lambda')$ indicate band index and $|k \lambda> $ are eigenstates of Dirac Hamiltonian, $H_D$.
The  generalized dressed Kubo response function, $\chi_{\sigma\tau}$ could be written as 
\begin{eqnarray}
\chi^{(\eta)}_{\sigma_z\tau_z}= \int \frac{d^2k}{(2\pi)^2} Tr[\tau_zG^{R(\eta)}(E_F,k) \sigma_z G^{A(\eta)}(E_F,k)]\nonumber\\
\end{eqnarray}

\section{Vertex correction}

The influence of impurities could be effectively formulated by considering different ways of interaction. First, independent interactions of electrons and hole with impurities. This effect can generally be captured by relaxation time calculated within the Born approximation. Then by replacing the Green's functions with  dressed ones as shown in the previous section the influence of independent impurities could be captured.  Second, by interaction of electron and hole pairs with a single impurity. This type of process appears as a set of pair interactions which could be shown as ladder type couplings depicted in Fig. 1. 
By means of a diagrammatic based concept named  \emph{vertex corrections} the effect of pair interactions can be included. 
 
As shown in Fig. 3, vertex corrections connect the electron and hole via the  interactions with impurities. Actually vertex correction is a parallel set of independent scatterings, so in the limit of ladder approximation, it leads to well-known Bethe-Salpeter self-consistence equations as \cite{27}
\begin{eqnarray}
\delta  \sigma_z= \bar{\sigma_z}+  n \displaystyle\sum_{k} V_{im}G^{R}_{k} \delta \sigma_z G^{A}_{k}V_{im}, 
\\
\bar{\sigma_z}= n \displaystyle\sum_{k} V_{im}G^{R}_{k} \sigma_z G^{A}_{k}V_{im}.
\end{eqnarray}
Where $n$ is the impurity density. By using the previous equations it can be shown that
\begin{eqnarray}
\delta  \sigma_z&=& n \displaystyle\sum_{k} V_{im}G^{R}_{k}(\sigma_z+\delta  \sigma_z)G^{A}_{k}V_{im}\\
&=&\frac{n}{(2\pi)^2}\int d^2k V_{im} G^{R}_{k}(\sigma_z+\delta  \sigma_z)G^{A}_{k}V_{im}. \nonumber
\end{eqnarray}

Finally, after substitution of $\sigma_z$ with $\sigma_z+\delta \sigma_z$  in (2.18), we can write the final expression for dressed response function under vertex corrections as 
\begin{eqnarray}
\chi_{\sigma_z\tau_z}= \int \frac{d^2k}{(2\pi)^2} Tr[\tau_z G^{R}_{k}(\sigma_z+\delta  \sigma_z)G^{A}_{k}].
\end{eqnarray}
\section{Results and Discussion}
Here, we present the numerical results which mainly contain $\chi^{\eta}_{\sigma_z\tau_z}$ response function in term of the Rashba coupling strength and normal field, $E_z$. The results have been obtained by numerical computations based on theoretical approach that  have been made clear in previous sections.

Figs. 4 and 5 show the behavior of  spin pseudo-spin response function $\chi^{\eta}_{\sigma_z\tau_z}$  in term of the extrinsic Rashba SOI coefficient $\lambda_{R1}$. 
\begin{figure}[h]
	\includegraphics[width=1.0\linewidth]{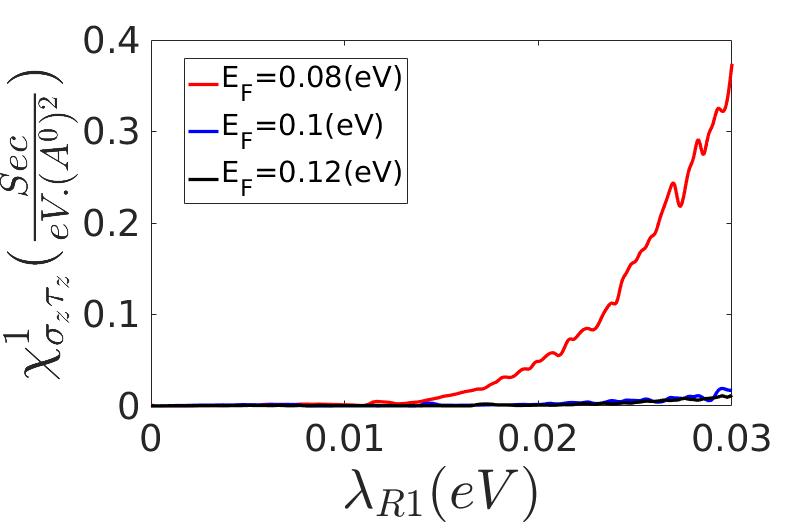}
	\centering
	\caption{(Color online)Spin pseudo-spin response of graphene function in term of the extrinsic Rashba coupling strength at $K$ point and different Fermi energies. \label{K_point_g}}
\end{figure}
As shown in these figures, it can be inferred that  increasing the Rashba coupling strength, $\lambda_{R1}$, leads to amplification of $\chi^{\eta}_{\sigma_z\tau_z}$  for both valleys. To provide a clear understanding about the contribution of intrinsic SOI, results have been compared with the case in which intrinsic SOI has been neglected for both valleys in graphene as shown in Figs. 6 and 7.
\\
\begin{figure}[h]
	\includegraphics[width=1.0\linewidth]{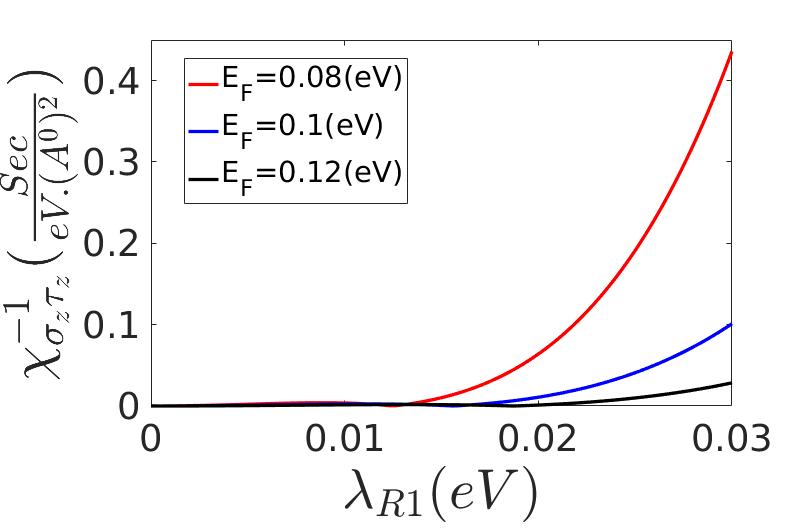}
	\centering
	\caption{(Color online) Spin pseudo-spin response function of graphene in term of the extrinsic Rashba coupling strength at $K'$ point and different Fermi energies.\label{Kp_point_g}}
\end{figure}

It can be identified from Figs. 6 and 7 that, the main source of spin to pseudo-spin conversion is the extrinsic Rashba interaction that establishes electronic-states in which the external magnetic field leads to pseudo-spin polarizing transitions. In the other words, the Rashba interaction changes the band states, $|k,\lambda>$, so that normal spin population imbalance gives rise
non-equilibrium pseudo-spin polarization. 
This interaction can also provide a framework for effective influence of intrinsic SOI on spin pseudo-spin response function. It should be noted that decreasing the Rashba coupling strength suppresses the contribution of intrinsic SOI and at the limit of $\lambda_{R1}\rightarrow 0$ spin pseudo-spin response function vanishes. This means that the intrinsic SOI has not an independent contribution in this effect and as it can be seen in the mentioned figures response function vanishes at the limit of zero Rashba coupling strength $\lambda_{R1}\rightarrow 0$.
\\

\begin{figure}[t]
	\includegraphics[width=1.0\linewidth]{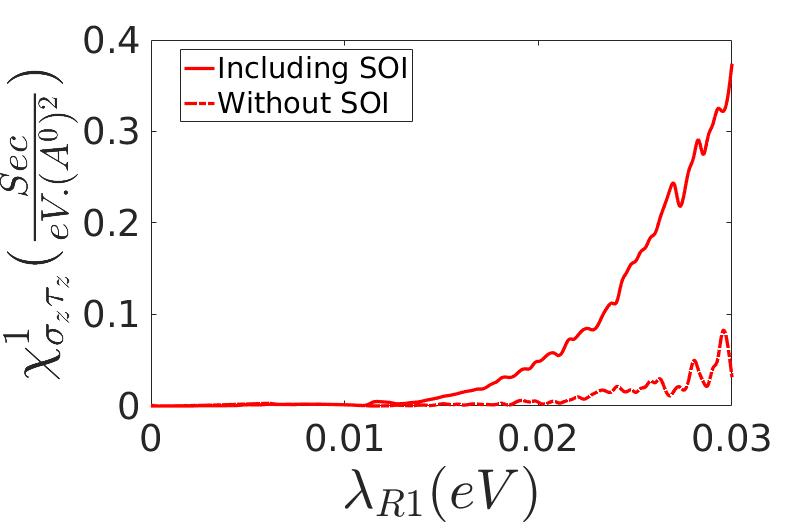}
	\centering
	\caption{(Color online) Spin to pseudo-spin conversion with (solid line) and without (dashed line) the SOI at $K$ point in graphene.\label{K_SIO_g}}
\end{figure}
\begin{figure}[h]
	\includegraphics[width=1.0\linewidth]{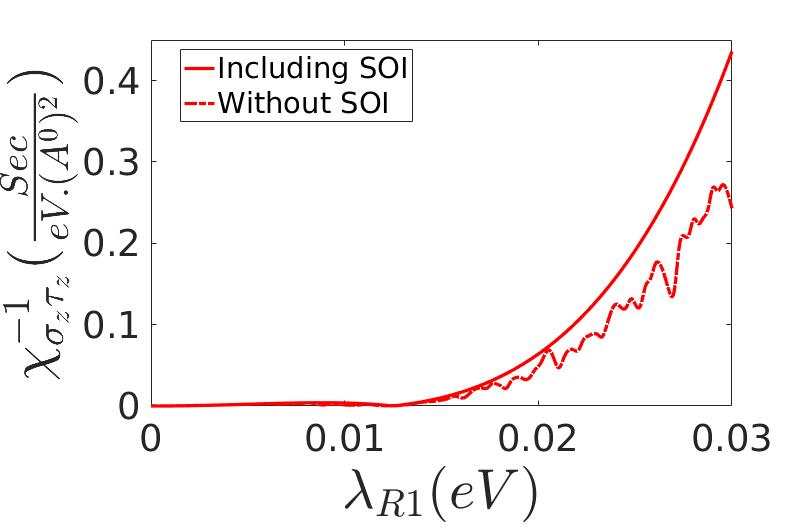}
	\centering
	\caption{(Color online) Spin to pseudo-spin conversion with (solid line) and without (dashed line) the SOI at $K'$ point in graphene. \label{Kp_SIO_g}}
\end{figure}
Response function of different valleys are not the same and a little difference could be observed. In other words, graphene's spin induced non-equilibrium pseudo-spin accompanied with a non-equilibrium valley polarization. This returns to the valley-dependent functional form of effective pseudo-magnetic field  defined in  Eq. (\ref{2.6}). This effective field leads to different non-equilibrium pseudo-spin polarization at these valleys which results small valley polarization ($\mathbf{P}$) which has been defined as
\begin{eqnarray}
P=\frac{\chi^{1}_{\sigma_{z}\tau_{z}}-\chi^{-1}_{\sigma_{z}\tau_{z}}}{\chi^{1}_{\sigma_{z}\tau_{z}}+\chi^{-1}_{\sigma_{z}\tau_{z}}}
\end{eqnarray} 
Meanwhile, numerical calculations show that valley dependence appears in a more pronounced manner  taking into account vertex corrections.
Referring to Eq. (\ref{2.16}) position independent part of the impurity scattering potential has the functional form exactly the same as intrinsic SOI.
Figs. 6 and 7 indicate the significant influence of vertex correction in increasing the valley asymmetry.  This can be explained if we consider that unlike the intrinsic SOI that changes its sign at different valleys, spin-orbit type potential of the impurities has a fix sign which means that the effective SOI of both lattice atoms and impurities, is not the same at each valleys. 

Furthermore, as it can be seen from Figs. 6 and 7, 
small oscillations of response function can be traced to the contribution of vertex corrections where 
in the absence of vertex corrections, $\chi^{\pm1}_{\sigma_z\tau_z}$  shows monotonic behavior as a function of the Rashba interaction. Therefore, the effect of successive scatterings which has been captured by vertex correction are
responsible for both small oscillations and increasing the valley dependence of response function in graphene.
\\
\begin{figure}[t]
	\includegraphics[width=1.0\linewidth]{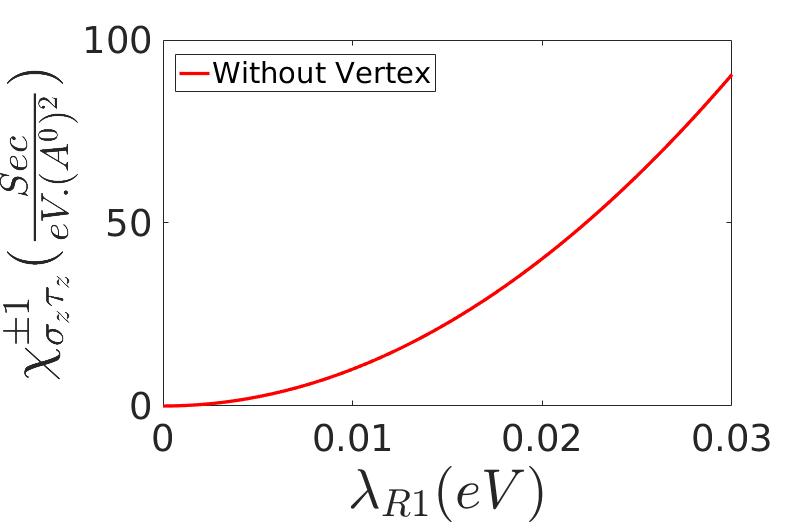}
	\centering
	\caption{(Color online) Spin to pseudo-spin conversion response function in term of the Rashba coupling strength in the absence of vertex corrections. The response function at both of the valleys are identical. \label{KKp_no_vertex_g}}
\end{figure}

In this way,  modulation of transition matrix, $<k|V_{im}|k'>$, due to the change of eigenstates by extrinsic Rashba may also lead to the small oscillations in $\chi^{\eta }_{\sigma_z\tau_z}$ in the case of graphene. 

It is very notable that vertex correction significantly decreases the spin to pseudo-spin response function for both valleys in graphene Fig. 8. In other words, stronger $\chi^{\pm 1}_{\sigma_z\tau_z}$can be obtained in the absence of relaxing mechanisms.

Fig. 9 shows the behavior of  spin pseudo-spin  response function $\chi^{\eta}_{\sigma_z\tau_z}$ in term of the external normal electric field, $E_z$, in silicene. As it can be recognized from this figure, it is obvious that the increasing of $E_z$ from $10(meV/ \AA)$ up to the critical electric field $E_c = 17(meV/ \AA)$, in which the band gap of silicene closes, causes an increase in  $\chi^{\pm 1}_{\sigma_z\tau_z}$. Actually it can be considered as a consequence of the  spin-valley locking and the effective pseudo-magnetic field amplification which gives rise to more pseudo-spin polarization. At the critical point $E_{c}$ i.e. at the gap closing electric filed, although  spin-valley locking reaches it's maximum value \cite{63}, however, the relaxation processes weaken the response function as a result of the elastic scatterings at this regime.  Meanwhile, increasing of $E_z$ beyond the critical value of $ 17(meV/ \AA)$ up to $ 20(meV/ \AA)$ reopens the gap therefore, relaxation rate of the elastic scatterings decreases which results in   abruptly increasing of response function. Meanwhile, increasing the normal electric field which acts as a pseudo magnetic field finally freezes the pseudo-spin quantum number of the electrons that leads to reduction of the response function at high electric fields. Exactly at the critical electric field response function sharply decreases.  
Therefore it seems that change of the band topology could be detected by this type of the response function. 

\begin{figure}[t]
	\includegraphics[width=1.0\linewidth]{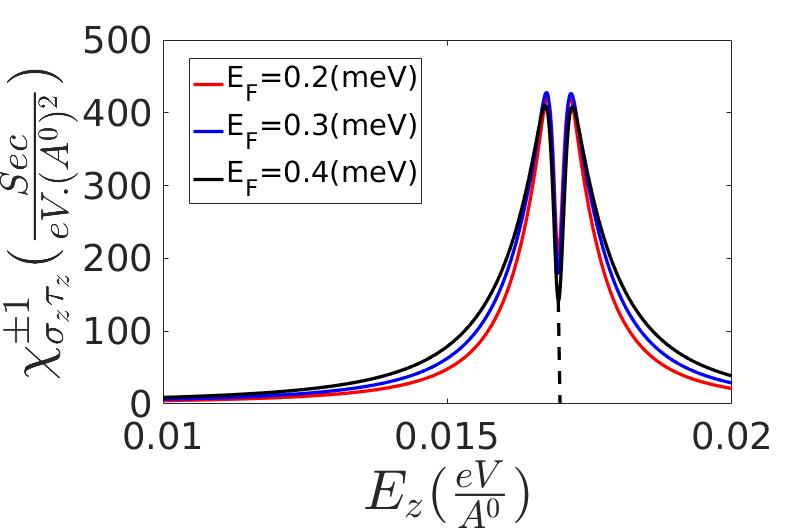}
	\centering
	\caption{(Color online) Spin pseudo-spin response function in term of the staggered electric field at different Fermi energies in silicene valleys. The dashed vertical line indicates the critical normal field of topological phase transition ($E_c$). \label{KKp_s}}
\end{figure}

Unlike the graphene, valley dependence of spin pseudo-spin  response function in  silicene is very small due to the order of magnitude difference in strength of valley independent term of the Hamiltonian, i.e. buckling term identified by $lE_z$, as compared to valley dependent SOIs. On the other hand strong k-dependent terms are suppressed linearly close to the Dirac points ($k\rightarrow 0$).  According to Eq. (\ref{2.7}) the buckling term is the only valley-independent and dominant term in the effective pseudo-magnetic field for silicene $B^{eff}_{Spm}$. In fact this term, identically contributes in spin pseudo-spin response functions for both of the valleys.

To recognizing the influence of scatterings and vertex correction on spin to pseudo-spin conversion in silicene, we have compared the obtained results with the case in which the contribution of vertex corrections have been omitted (Fig. 10). 
According to these results, interestingly it has been found that in contrast to the graphene, vertex correction has strengthened the response function of both valleys in the case of silicene. To explain this effect one should consider nature of the impurities chosen in this work and different contribution of scatterings in gap-less and gapped Dirac materials.
\\
\begin{figure}[t]
	\includegraphics[width=1.0\linewidth]{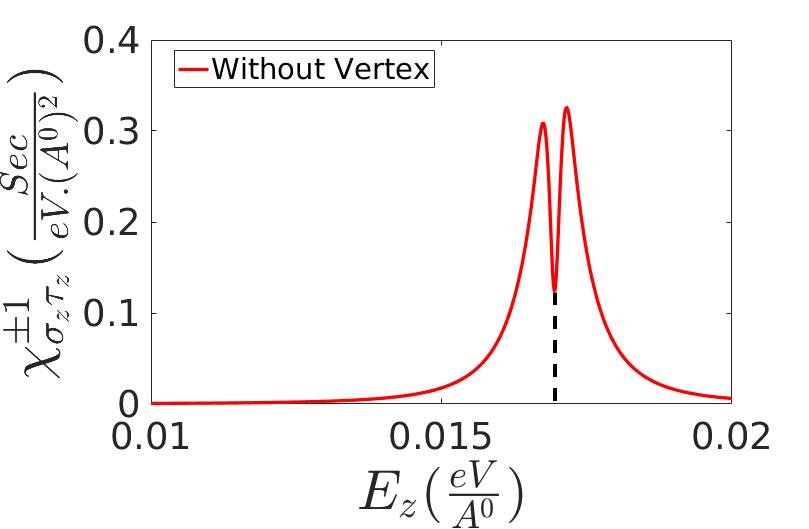}
	\centering
	\caption{(Color online) Spin pseudo-spin response function of silicene in term of the staggered electric field at each of the valleys when vertex corrections have been ignored. The dashed vertical line indicates the critical gap closing electric field. \label{eps-mg-0-2d}}
\end{figure}
In the presence of 
buckling and applied normal electric field, $E_z$, it can be shown that normal spin and pseudo-spin are good quantum numbers since exactly at the Dirac point, Hamiltonian commutes with these operators. Very close to the Dirac points $k \rightarrow 0$ since the Rashba coupling strength is very small in silicene, one can see that the Hamiltonian (Eq. (\ref{Hamiltonian})) is reduced to $H^{\eta}_D \approx \eta\lambda _{so}  \tau _z\sigma _z -l E_z \tau _z$. Therefore, each of the bands has a definite pseudo-spin index i.e. very close to the Dirac points each of bands is almost pseudo-spin polarized. The polarization range in $k$-space depends on the magnitude of $E_z$. I In the other words, spin polarized valleys as previously reported \cite{63} leads to pseudo-spin polarized bands. Therefore, all of the bands has its own specific $\tau_z$ and spin-valley locking in silicene leads to invariant $\sigma_z$ at each of the valleys. Due to this fact inter-band scatterings could relax the pseudo-spin polarization and inter-valley transitions could result in spin relaxation.  Unlike the momentum operator, $\sigma_z$ and $\tau_z$ commute with impurity potential which means that impurities should preserve the spin and pseudo-spin at band polarized regime ($E_z \neq 0$) in silicene. This protects each of the excited states from relaxation at insulating regime when the elastic scatterings are absent. Accordingly, it can be realized that the vertex correction which takes into account the successive interactions of electron-hole pair of different bands with impurities cannot contribute into the relaxation of the non-equilibrium normal spins. Meanwhile, elastic scatterings are less effective in the insulating regime. Furthermore the vertex correction, gives rise to stronger normalized motive $\sigma_z$ in the case of gapped silicene which results in enhancement of the response function.

By contrast, vertex correction resonates scatterings in graphene, because pseudo-spin is not good quantum number to be protected by spin and pseudo-spin preserving scatterings. This is due to the fact that unlike silicene the intrinsic SOI in graphene is small in comparison with Rashba coupling strength. Accordingly, the effective Hamiltonian of the graphene very close to the Dirac point in the absence of buckling is $H^{\eta}_D \approx  \frac{1}{2} \lambda _{R1} \left(\eta  \tau _x \sigma
_y-\sigma _x \tau _y\right).$  In this case where there is no pseudo-spin polarized band and no spin polarized valley, impurities with spin-orbit type couplings can effectively contribute in the band mixing and therefore in the pseudo-spin relaxation.
 
It seems that the use Dirac point approximation in which the inter-valley transitions have been ignored, results in protection of normal spins against the relaxations that could be made by inter-valley scatterings in a non-physical manner. Inter-valley intra-band transitions can be provided by short range and sharp-varying potential of impurities. However, it should be noted that even when the calculations are performed beyond the Dirac point approximation due to the well defined opposite normal spin of valleys at low Fermi energies, inter-valley transitions cannot take place by an impurity potential that commutes with $\sigma_{z}$ in this limit. Therefor, spin-valley locking which leads to spin-resolved valleys could prevent the normal spin relaxations in the presence of non-magnetic or $\sigma_{z}$-commuting impurity potentials.

In-addition, pseudo-spin polarized bands in silicene provided by normally applied field and structural inversion asymmetry (buckling) may bring in mind what the need for magnetically generated pseudo-spin polarization when the system polarized at equilibrium. It should be noted that in the present work non-equilibrium pseudo-spin polarization has been obtained by real magnetic field and response functions just measure quantities generated at non-equilibrium regime as a result of the external perturbations. Meanwhile via the Onsager relarions non-equilibrium pseudo-spin could result in non-equilibrium spin polarizations. Therefore, this approach provides a practical way through magnetic detection of non-equilibrium pseudo-spin.

For Dirac materials in metallic phase such as graphene or in the case of gap closing silicene at the critical electric field, elastic scatterings can contribute in the pseudo-spin relaxation. This elastic type of relaxations is identified by the Dirac delta functions in Eq. (\ref{2.17}) that provides a definite relaxation time for each of states. Elastic scatterings could take place via the intra-band  or even inter-band transitions (if the Fermi level is located at band width overlap of different bands and when the bands are not pseudo-spin resolved).  The influence of elastic relaxations have been taken into account in the dressed Green's functions using the Born approximation. These type of relaxations could contribute in the reduction of polarizations in metallic graphene or can explain the response function fall of silicene at the critical electric field (Fig. 9).  

\section{Conclusion} 
A novel type of conversion called spin to pseudo-spin has been proposed which originally arises from the Rashba and spin-orbit interactions for graphene-like materials. In this way, the spin pseudo-spin response function, $\chi^{\eta}_{\sigma_z\tau_z}$, has been computed in the context of Kubo formalism by taking into account vertex corrections. In the case of graphene, The obtained response function increases by increasing the extrinsic Rashba coupling strength ($\lambda_{R1}$). Meanwhile, valley dependence of the response function in graphene comes from the functional form of position independent part of scattering matrix ($M$) which in the present case has been chosen same as intrinsic SOI.

In the case of silicene, spin pseudo-spin response function  has been obtained in the presence of a normally applied electric field. At typical values of this electric field and close to the Dirac points contribution of this field is dominant  compared with all types the SOIs. Accordingly, since the electric field interaction has the same contribution at each of the valleys, therefore the valley dependence of the response function has been suppressed by the normal electric field.

Unlike the graphene, vertex correction enhances the spin pseudo-spin response function in the  silicene structure.  Spin-valley locking and pseudo-spin polarized bands, that appear as a result of the normal electric field in buckled silicene, make the spin and pseudo-spin as vertex protected quantum numbers in the presence of spin-orbit type impurities. This leads to significant enhancement of the response function of silicene. It can also be inferred that the real-space dependent part of the impurity potential could be effective on momentum relaxation and cannot result in inter-band transitions between the pseudo-spin resolved bands.

\bibliographystyle{apsrev4-1}
\bibliography{refrences}

\begin{thebibliography}{65}%
\makeatletter
\providecommand \@ifxundefined [1]{%
 \@ifx{#1\undefined}
}%
\providecommand \@ifnum [1]{%
 \ifnum #1\expandafter \@firstoftwo
 \else \expandafter \@secondoftwo
 \fi
}%
\providecommand \@ifx [1]{%
 \ifx #1\expandafter \@firstoftwo
 \else \expandafter \@secondoftwo
 \fi
}%
\providecommand \natexlab [1]{#1}%
\providecommand \enquote  [1]{``#1''}%
\providecommand \bibnamefont  [1]{#1}%
\providecommand \bibfnamefont [1]{#1}%
\providecommand \citenamefont [1]{#1}%
\providecommand \href@noop [0]{\@secondoftwo}%
\providecommand \href [0]{\begingroup \@sanitize@url \@href}%
\providecommand \@href[1]{\@@startlink{#1}\@@href}%
\providecommand \@@href[1]{\endgroup#1\@@endlink}%
\providecommand \@sanitize@url [0]{\catcode `\\12\catcode `\$12\catcode
  `\&12\catcode `\#12\catcode `\^12\catcode `\_12\catcode `\%12\relax}%
\providecommand \@@startlink[1]{}%
\providecommand \@@endlink[0]{}%
\providecommand \url  [0]{\begingroup\@sanitize@url \@url }%
\providecommand \@url [1]{\endgroup\@href {#1}{\urlprefix }}%
\providecommand \urlprefix  [0]{URL }%
\providecommand \Eprint [0]{\href }%
\providecommand \doibase [0]{http://dx.doi.org/}%
\providecommand \selectlanguage [0]{\@gobble}%
\providecommand \bibinfo  [0]{\@secondoftwo}%
\providecommand \bibfield  [0]{\@secondoftwo}%
\providecommand \translation [1]{[#1]}%
\providecommand \BibitemOpen [0]{}%
\providecommand \bibitemStop [0]{}%
\providecommand \bibitemNoStop [0]{.\EOS\space}%
\providecommand \EOS [0]{\spacefactor3000\relax}%
\providecommand \BibitemShut  [1]{\csname bibitem#1\endcsname}%
\let\auto@bib@innerbib\@empty
\bibitem [{\citenamefont {Wolf}\ \emph {et~al.}(2001)\citenamefont {Wolf},
  \citenamefont {Awschalom}, \citenamefont {Buhrman}, \citenamefont {Daughton},
  \citenamefont {Von~Molnar}, \citenamefont {Roukes}, \citenamefont
  {Chtchelkanova},\ and\ \citenamefont {Treger}}]{1}%
  \BibitemOpen
  \bibfield  {author} {\bibinfo {author} {\bibfnamefont {S.}~\bibnamefont
  {Wolf}}, \bibinfo {author} {\bibfnamefont {D.}~\bibnamefont {Awschalom}},
  \bibinfo {author} {\bibfnamefont {R.}~\bibnamefont {Buhrman}}, \bibinfo
  {author} {\bibfnamefont {J.}~\bibnamefont {Daughton}}, \bibinfo {author}
  {\bibfnamefont {S.}~\bibnamefont {Von~Molnar}}, \bibinfo {author}
  {\bibfnamefont {M.}~\bibnamefont {Roukes}}, \bibinfo {author} {\bibfnamefont
  {A.~Y.}\ \bibnamefont {Chtchelkanova}}, \ and\ \bibinfo {author}
  {\bibfnamefont {D.}~\bibnamefont {Treger}},\ }\href@noop {} {\bibfield
  {journal} {\bibinfo  {journal} {Science}\ }\textbf {\bibinfo {volume}
  {294}},\ \bibinfo {pages} {1488} (\bibinfo {year} {2001})}\BibitemShut
  {NoStop}%
\bibitem [{\citenamefont {Nikonov}\ \emph {et~al.}(2006)\citenamefont
  {Nikonov}, \citenamefont {Bourianoff},\ and\ \citenamefont {Gargini}}]{2}%
  \BibitemOpen
  \bibfield  {author} {\bibinfo {author} {\bibfnamefont {D.~E.}\ \bibnamefont
  {Nikonov}}, \bibinfo {author} {\bibfnamefont {G.~I.}\ \bibnamefont
  {Bourianoff}}, \ and\ \bibinfo {author} {\bibfnamefont {P.~A.}\ \bibnamefont
  {Gargini}},\ }\href@noop {} {\bibfield  {journal} {\bibinfo  {journal}
  {Journal of superconductivity and novel magnetism}\ }\textbf {\bibinfo
  {volume} {19}},\ \bibinfo {pages} {497} (\bibinfo {year} {2006})}\BibitemShut
  {NoStop}%
\bibitem [{\citenamefont {{\v{Z}}uti{\'c}}\ \emph {et~al.}(2004)\citenamefont
  {{\v{Z}}uti{\'c}}, \citenamefont {Fabian},\ and\ \citenamefont {Sarma}}]{3}%
  \BibitemOpen
  \bibfield  {author} {\bibinfo {author} {\bibfnamefont {I.}~\bibnamefont
  {{\v{Z}}uti{\'c}}}, \bibinfo {author} {\bibfnamefont {J.}~\bibnamefont
  {Fabian}}, \ and\ \bibinfo {author} {\bibfnamefont {S.~D.}\ \bibnamefont
  {Sarma}},\ }\href@noop {} {\bibfield  {journal} {\bibinfo  {journal} {Reviews
  of modern physics}\ }\textbf {\bibinfo {volume} {76}},\ \bibinfo {pages}
  {323} (\bibinfo {year} {2004})}\BibitemShut {NoStop}%
\bibitem [{\citenamefont {Sinova}\ \emph {et~al.}(2015)\citenamefont {Sinova},
  \citenamefont {Valenzuela}, \citenamefont {Wunderlich}, \citenamefont
  {Back},\ and\ \citenamefont {Jungwirth}}]{4}%
  \BibitemOpen
  \bibfield  {author} {\bibinfo {author} {\bibfnamefont {J.}~\bibnamefont
  {Sinova}}, \bibinfo {author} {\bibfnamefont {S.~O.}\ \bibnamefont
  {Valenzuela}}, \bibinfo {author} {\bibfnamefont {J.}~\bibnamefont
  {Wunderlich}}, \bibinfo {author} {\bibfnamefont {C.}~\bibnamefont {Back}}, \
  and\ \bibinfo {author} {\bibfnamefont {T.}~\bibnamefont {Jungwirth}},\
  }\href@noop {} {\bibfield  {journal} {\bibinfo  {journal} {Reviews of Modern
  Physics}\ }\textbf {\bibinfo {volume} {87}},\ \bibinfo {pages} {1213}
  (\bibinfo {year} {2015})}\BibitemShut {NoStop}%
\bibitem [{\citenamefont {Van~Tuan}\ and\ \citenamefont {Roche}(2016)}]{51}%
  \BibitemOpen
  \bibfield  {author} {\bibinfo {author} {\bibfnamefont {D.}~\bibnamefont
  {Van~Tuan}}\ and\ \bibinfo {author} {\bibfnamefont {S.}~\bibnamefont
  {Roche}},\ }\href@noop {} {\bibfield  {journal} {\bibinfo  {journal}
  {Physical review letters}\ }\textbf {\bibinfo {volume} {116}},\ \bibinfo
  {pages} {106601} (\bibinfo {year} {2016})}\BibitemShut {NoStop}%
\bibitem [{\citenamefont {Zhang}(2000)}]{5}%
  \BibitemOpen
  \bibfield  {author} {\bibinfo {author} {\bibfnamefont {S.}~\bibnamefont
  {Zhang}},\ }\href@noop {} {\bibfield  {journal} {\bibinfo  {journal}
  {Physical review letters}\ }\textbf {\bibinfo {volume} {85}},\ \bibinfo
  {pages} {393} (\bibinfo {year} {2000})}\BibitemShut {NoStop}%
\bibitem [{\citenamefont {Kato}\ \emph {et~al.}(2004)\citenamefont {Kato},
  \citenamefont {Myers}, \citenamefont {Gossard},\ and\ \citenamefont
  {Awschalom}}]{6}%
  \BibitemOpen
  \bibfield  {author} {\bibinfo {author} {\bibfnamefont {Y.~K.}\ \bibnamefont
  {Kato}}, \bibinfo {author} {\bibfnamefont {R.~C.}\ \bibnamefont {Myers}},
  \bibinfo {author} {\bibfnamefont {A.~C.}\ \bibnamefont {Gossard}}, \ and\
  \bibinfo {author} {\bibfnamefont {D.~D.}\ \bibnamefont {Awschalom}},\
  }\href@noop {} {\bibfield  {journal} {\bibinfo  {journal} {science}\ }\textbf
  {\bibinfo {volume} {306}},\ \bibinfo {pages} {1910} (\bibinfo {year}
  {2004})}\BibitemShut {NoStop}%
\bibitem [{\citenamefont {Jungwirth}\ \emph {et~al.}()\citenamefont
  {Jungwirth}, \citenamefont {Wunderlich},\ and\ \citenamefont {Olejnik}}]{7}%
  \BibitemOpen
  \bibfield  {author} {\bibinfo {author} {\bibfnamefont {T.}~\bibnamefont
  {Jungwirth}}, \bibinfo {author} {\bibfnamefont {J.}~\bibnamefont
  {Wunderlich}}, \ and\ \bibinfo {author} {\bibfnamefont {K.}~\bibnamefont
  {Olejnik}},\ }\href@noop {} {\enquote {\bibinfo {title} {Nature mater. 11,
  382 (2012)},}\ }\BibitemShut {NoStop}%
\bibitem [{\citenamefont {Hoffmann}(2013)}]{8}%
  \BibitemOpen
  \bibfield  {author} {\bibinfo {author} {\bibfnamefont {A.}~\bibnamefont
  {Hoffmann}},\ }\href@noop {} {\bibfield  {journal} {\bibinfo  {journal} {IEEE
  transactions on magnetics}\ }\textbf {\bibinfo {volume} {49}},\ \bibinfo
  {pages} {5172} (\bibinfo {year} {2013})}\BibitemShut {NoStop}%
\bibitem [{\citenamefont {Murakami}\ \emph {et~al.}(2003)\citenamefont
  {Murakami}, \citenamefont {Nagaosa},\ and\ \citenamefont {Zhang}}]{9}%
  \BibitemOpen
  \bibfield  {author} {\bibinfo {author} {\bibfnamefont {S.}~\bibnamefont
  {Murakami}}, \bibinfo {author} {\bibfnamefont {N.}~\bibnamefont {Nagaosa}}, \
  and\ \bibinfo {author} {\bibfnamefont {S.-C.}\ \bibnamefont {Zhang}},\
  }\href@noop {} {\bibfield  {journal} {\bibinfo  {journal} {Science}\ }\textbf
  {\bibinfo {volume} {301}},\ \bibinfo {pages} {1348} (\bibinfo {year}
  {2003})}\BibitemShut {NoStop}%
\bibitem [{\citenamefont {Maekawa}\ \emph {et~al.}(2013)\citenamefont
  {Maekawa}, \citenamefont {Adachi}, \citenamefont {Uchida}, \citenamefont
  {Ieda},\ and\ \citenamefont {Saitoh}}]{10}%
  \BibitemOpen
  \bibfield  {author} {\bibinfo {author} {\bibfnamefont {S.}~\bibnamefont
  {Maekawa}}, \bibinfo {author} {\bibfnamefont {H.}~\bibnamefont {Adachi}},
  \bibinfo {author} {\bibfnamefont {K.-i.}\ \bibnamefont {Uchida}}, \bibinfo
  {author} {\bibfnamefont {J.}~\bibnamefont {Ieda}}, \ and\ \bibinfo {author}
  {\bibfnamefont {E.}~\bibnamefont {Saitoh}},\ }\href@noop {} {\bibfield
  {journal} {\bibinfo  {journal} {Journal of the Physical Society of Japan}\
  }\textbf {\bibinfo {volume} {82}},\ \bibinfo {pages} {102002} (\bibinfo
  {year} {2013})}\BibitemShut {NoStop}%
\bibitem [{\citenamefont {Manchon}\ \emph {et~al.}(2015)\citenamefont
  {Manchon}, \citenamefont {Koo}, \citenamefont {Nitta}, \citenamefont
  {Frolov},\ and\ \citenamefont {Duine}}]{11}%
  \BibitemOpen
  \bibfield  {author} {\bibinfo {author} {\bibfnamefont {A.}~\bibnamefont
  {Manchon}}, \bibinfo {author} {\bibfnamefont {H.~C.}\ \bibnamefont {Koo}},
  \bibinfo {author} {\bibfnamefont {J.}~\bibnamefont {Nitta}}, \bibinfo
  {author} {\bibfnamefont {S.}~\bibnamefont {Frolov}}, \ and\ \bibinfo {author}
  {\bibfnamefont {R.}~\bibnamefont {Duine}},\ }\href@noop {} {\bibfield
  {journal} {\bibinfo  {journal} {Nature materials}\ }\textbf {\bibinfo
  {volume} {14}},\ \bibinfo {pages} {871} (\bibinfo {year} {2015})}\BibitemShut
  {NoStop}%
\bibitem [{\citenamefont {Ando}\ and\ \citenamefont {Shiraishi}(2016)}]{12}%
  \BibitemOpen
  \bibfield  {author} {\bibinfo {author} {\bibfnamefont {Y.}~\bibnamefont
  {Ando}}\ and\ \bibinfo {author} {\bibfnamefont {M.}~\bibnamefont
  {Shiraishi}},\ }\href@noop {} {\bibfield  {journal} {\bibinfo  {journal}
  {Journal of the Physical Society of Japan}\ }\textbf {\bibinfo {volume}
  {86}},\ \bibinfo {pages} {011001} (\bibinfo {year} {2016})}\BibitemShut
  {NoStop}%
\bibitem [{\citenamefont {Han}\ \emph {et~al.}(2018)\citenamefont {Han},
  \citenamefont {Otani},\ and\ \citenamefont {Maekawa}}]{13}%
  \BibitemOpen
  \bibfield  {author} {\bibinfo {author} {\bibfnamefont {W.}~\bibnamefont
  {Han}}, \bibinfo {author} {\bibfnamefont {Y.}~\bibnamefont {Otani}}, \ and\
  \bibinfo {author} {\bibfnamefont {S.}~\bibnamefont {Maekawa}},\ }\href@noop
  {} {\bibfield  {journal} {\bibinfo  {journal} {npj Quantum Materials}\
  }\textbf {\bibinfo {volume} {3}},\ \bibinfo {pages} {27} (\bibinfo {year}
  {2018})}\BibitemShut {NoStop}%
\bibitem [{\citenamefont {S{\'a}nchez}\ \emph {et~al.}(2013)\citenamefont
  {S{\'a}nchez}, \citenamefont {Vila}, \citenamefont {Desfonds}, \citenamefont
  {Gambarelli}, \citenamefont {Attan{\'e}}, \citenamefont {De~Teresa},
  \citenamefont {Mag{\'e}n},\ and\ \citenamefont {Fert}}]{14}%
  \BibitemOpen
  \bibfield  {author} {\bibinfo {author} {\bibfnamefont {J.~R.}\ \bibnamefont
  {S{\'a}nchez}}, \bibinfo {author} {\bibfnamefont {L.}~\bibnamefont {Vila}},
  \bibinfo {author} {\bibfnamefont {G.}~\bibnamefont {Desfonds}}, \bibinfo
  {author} {\bibfnamefont {S.}~\bibnamefont {Gambarelli}}, \bibinfo {author}
  {\bibfnamefont {J.}~\bibnamefont {Attan{\'e}}}, \bibinfo {author}
  {\bibfnamefont {J.}~\bibnamefont {De~Teresa}}, \bibinfo {author}
  {\bibfnamefont {C.}~\bibnamefont {Mag{\'e}n}}, \ and\ \bibinfo {author}
  {\bibfnamefont {A.}~\bibnamefont {Fert}},\ }\href@noop {} {\bibfield
  {journal} {\bibinfo  {journal} {Nature communications}\ }\textbf {\bibinfo
  {volume} {4}},\ \bibinfo {pages} {2944} (\bibinfo {year} {2013})}\BibitemShut
  {NoStop}%
\bibitem [{\citenamefont {Soumyanarayanan}\ \emph {et~al.}(2016)\citenamefont
  {Soumyanarayanan}, \citenamefont {Reyren}, \citenamefont {Fert},\ and\
  \citenamefont {Panagopoulos}}]{15}%
  \BibitemOpen
  \bibfield  {author} {\bibinfo {author} {\bibfnamefont {A.}~\bibnamefont
  {Soumyanarayanan}}, \bibinfo {author} {\bibfnamefont {N.}~\bibnamefont
  {Reyren}}, \bibinfo {author} {\bibfnamefont {A.}~\bibnamefont {Fert}}, \ and\
  \bibinfo {author} {\bibfnamefont {C.}~\bibnamefont {Panagopoulos}},\
  }\href@noop {} {\bibfield  {journal} {\bibinfo  {journal} {Nature}\ }\textbf
  {\bibinfo {volume} {539}},\ \bibinfo {pages} {509} (\bibinfo {year}
  {2016})}\BibitemShut {NoStop}%
\bibitem [{\citenamefont {Seibold}\ \emph {et~al.}(2017)\citenamefont
  {Seibold}, \citenamefont {Caprara}, \citenamefont {Grilli},\ and\
  \citenamefont {Raimondi}}]{16}%
  \BibitemOpen
  \bibfield  {author} {\bibinfo {author} {\bibfnamefont {G.}~\bibnamefont
  {Seibold}}, \bibinfo {author} {\bibfnamefont {S.}~\bibnamefont {Caprara}},
  \bibinfo {author} {\bibfnamefont {M.}~\bibnamefont {Grilli}}, \ and\ \bibinfo
  {author} {\bibfnamefont {R.}~\bibnamefont {Raimondi}},\ }\href@noop {}
  {\bibfield  {journal} {\bibinfo  {journal} {Physical review letters}\
  }\textbf {\bibinfo {volume} {119}},\ \bibinfo {pages} {256801} (\bibinfo
  {year} {2017})}\BibitemShut {NoStop}%
\bibitem [{\citenamefont {Dyakonov}\ and\ \citenamefont {Perel}(1971)}]{17}%
  \BibitemOpen
  \bibfield  {author} {\bibinfo {author} {\bibfnamefont {M.}~\bibnamefont
  {Dyakonov}}\ and\ \bibinfo {author} {\bibfnamefont {V.}~\bibnamefont
  {Perel}},\ }\href@noop {} {\bibfield  {journal} {\bibinfo  {journal} {Physics
  Letters A}\ }\textbf {\bibinfo {volume} {35}},\ \bibinfo {pages} {459}
  (\bibinfo {year} {1971})}\BibitemShut {NoStop}%
\bibitem [{\citenamefont {Hirsch}(1999)}]{18}%
  \BibitemOpen
  \bibfield  {author} {\bibinfo {author} {\bibfnamefont {J.}~\bibnamefont
  {Hirsch}},\ }\href@noop {} {\bibfield  {journal} {\bibinfo  {journal}
  {Physical Review Letters}\ }\textbf {\bibinfo {volume} {83}},\ \bibinfo
  {pages} {1834} (\bibinfo {year} {1999})}\BibitemShut {NoStop}%
\bibitem [{\citenamefont {Mosendz}\ \emph {et~al.}(2010)\citenamefont
  {Mosendz}, \citenamefont {Vlaminck}, \citenamefont {Pearson}, \citenamefont
  {Fradin}, \citenamefont {Bauer}, \citenamefont {Bader},\ and\ \citenamefont
  {Hoffmann}}]{19}%
  \BibitemOpen
  \bibfield  {author} {\bibinfo {author} {\bibfnamefont {O.}~\bibnamefont
  {Mosendz}}, \bibinfo {author} {\bibfnamefont {V.}~\bibnamefont {Vlaminck}},
  \bibinfo {author} {\bibfnamefont {J.}~\bibnamefont {Pearson}}, \bibinfo
  {author} {\bibfnamefont {F.}~\bibnamefont {Fradin}}, \bibinfo {author}
  {\bibfnamefont {G.}~\bibnamefont {Bauer}}, \bibinfo {author} {\bibfnamefont
  {S.}~\bibnamefont {Bader}}, \ and\ \bibinfo {author} {\bibfnamefont
  {A.}~\bibnamefont {Hoffmann}},\ }\href@noop {} {\bibfield  {journal}
  {\bibinfo  {journal} {Physical Review B}\ }\textbf {\bibinfo {volume} {82}},\
  \bibinfo {pages} {214403} (\bibinfo {year} {2010})}\BibitemShut {NoStop}%
\bibitem [{\citenamefont {Tao}\ \emph {et~al.}(2018)\citenamefont {Tao},
  \citenamefont {Liu}, \citenamefont {Miao}, \citenamefont {Yu}, \citenamefont
  {Feng}, \citenamefont {Sun}, \citenamefont {You}, \citenamefont {Du},
  \citenamefont {Chen}, \citenamefont {Zhang} \emph {et~al.}}]{20}%
  \BibitemOpen
  \bibfield  {author} {\bibinfo {author} {\bibfnamefont {X.}~\bibnamefont
  {Tao}}, \bibinfo {author} {\bibfnamefont {Q.}~\bibnamefont {Liu}}, \bibinfo
  {author} {\bibfnamefont {B.}~\bibnamefont {Miao}}, \bibinfo {author}
  {\bibfnamefont {R.}~\bibnamefont {Yu}}, \bibinfo {author} {\bibfnamefont
  {Z.}~\bibnamefont {Feng}}, \bibinfo {author} {\bibfnamefont {L.}~\bibnamefont
  {Sun}}, \bibinfo {author} {\bibfnamefont {B.}~\bibnamefont {You}}, \bibinfo
  {author} {\bibfnamefont {J.}~\bibnamefont {Du}}, \bibinfo {author}
  {\bibfnamefont {K.}~\bibnamefont {Chen}}, \bibinfo {author} {\bibfnamefont
  {S.}~\bibnamefont {Zhang}},  \emph {et~al.},\ }\href@noop {} {\bibfield
  {journal} {\bibinfo  {journal} {Science advances}\ }\textbf {\bibinfo
  {volume} {4}},\ \bibinfo {pages} {eaat1670} (\bibinfo {year}
  {2018})}\BibitemShut {NoStop}%
\bibitem [{\citenamefont {Morota}\ \emph {et~al.}(2011)\citenamefont {Morota},
  \citenamefont {Niimi}, \citenamefont {Ohnishi}, \citenamefont {Wei},
  \citenamefont {Tanaka}, \citenamefont {Kontani}, \citenamefont {Kimura},\
  and\ \citenamefont {Otani}}]{21}%
  \BibitemOpen
  \bibfield  {author} {\bibinfo {author} {\bibfnamefont {M.}~\bibnamefont
  {Morota}}, \bibinfo {author} {\bibfnamefont {Y.}~\bibnamefont {Niimi}},
  \bibinfo {author} {\bibfnamefont {K.}~\bibnamefont {Ohnishi}}, \bibinfo
  {author} {\bibfnamefont {D.}~\bibnamefont {Wei}}, \bibinfo {author}
  {\bibfnamefont {T.}~\bibnamefont {Tanaka}}, \bibinfo {author} {\bibfnamefont
  {H.}~\bibnamefont {Kontani}}, \bibinfo {author} {\bibfnamefont
  {T.}~\bibnamefont {Kimura}}, \ and\ \bibinfo {author} {\bibfnamefont
  {Y.}~\bibnamefont {Otani}},\ }\href@noop {} {\bibfield  {journal} {\bibinfo
  {journal} {Physical Review B}\ }\textbf {\bibinfo {volume} {83}},\ \bibinfo
  {pages} {174405} (\bibinfo {year} {2011})}\BibitemShut {NoStop}%
\bibitem [{\citenamefont {Wunderlich}\ \emph {et~al.}(2005)\citenamefont
  {Wunderlich}, \citenamefont {Kaestner}, \citenamefont {Sinova},\ and\
  \citenamefont {Jungwirth}}]{22}%
  \BibitemOpen
  \bibfield  {author} {\bibinfo {author} {\bibfnamefont {J.}~\bibnamefont
  {Wunderlich}}, \bibinfo {author} {\bibfnamefont {B.}~\bibnamefont
  {Kaestner}}, \bibinfo {author} {\bibfnamefont {J.}~\bibnamefont {Sinova}}, \
  and\ \bibinfo {author} {\bibfnamefont {T.}~\bibnamefont {Jungwirth}},\
  }\href@noop {} {\bibfield  {journal} {\bibinfo  {journal} {Physical review
  letters}\ }\textbf {\bibinfo {volume} {94}},\ \bibinfo {pages} {047204}
  (\bibinfo {year} {2005})}\BibitemShut {NoStop}%
\bibitem [{\citenamefont {Liu}\ \emph {et~al.}(2012)\citenamefont {Liu},
  \citenamefont {Pai}, \citenamefont {Li}, \citenamefont {Tseng}, \citenamefont
  {Ralph},\ and\ \citenamefont {Buhrman}}]{23}%
  \BibitemOpen
  \bibfield  {author} {\bibinfo {author} {\bibfnamefont {L.}~\bibnamefont
  {Liu}}, \bibinfo {author} {\bibfnamefont {C.-F.}\ \bibnamefont {Pai}},
  \bibinfo {author} {\bibfnamefont {Y.}~\bibnamefont {Li}}, \bibinfo {author}
  {\bibfnamefont {H.}~\bibnamefont {Tseng}}, \bibinfo {author} {\bibfnamefont
  {D.}~\bibnamefont {Ralph}}, \ and\ \bibinfo {author} {\bibfnamefont
  {R.}~\bibnamefont {Buhrman}},\ }\href@noop {} {\bibfield  {journal} {\bibinfo
   {journal} {Science}\ }\textbf {\bibinfo {volume} {336}},\ \bibinfo {pages}
  {555} (\bibinfo {year} {2012})}\BibitemShut {NoStop}%
\bibitem [{\citenamefont {Wakamura}\ \emph {et~al.}(2015)\citenamefont
  {Wakamura}, \citenamefont {Akaike}, \citenamefont {Omori}, \citenamefont
  {Niimi}, \citenamefont {Takahashi}, \citenamefont {Fujimaki}, \citenamefont
  {Maekawa},\ and\ \citenamefont {Otani}}]{24}%
  \BibitemOpen
  \bibfield  {author} {\bibinfo {author} {\bibfnamefont {T.}~\bibnamefont
  {Wakamura}}, \bibinfo {author} {\bibfnamefont {H.}~\bibnamefont {Akaike}},
  \bibinfo {author} {\bibfnamefont {Y.}~\bibnamefont {Omori}}, \bibinfo
  {author} {\bibfnamefont {Y.}~\bibnamefont {Niimi}}, \bibinfo {author}
  {\bibfnamefont {S.}~\bibnamefont {Takahashi}}, \bibinfo {author}
  {\bibfnamefont {A.}~\bibnamefont {Fujimaki}}, \bibinfo {author}
  {\bibfnamefont {S.}~\bibnamefont {Maekawa}}, \ and\ \bibinfo {author}
  {\bibfnamefont {Y.}~\bibnamefont {Otani}},\ }\href@noop {} {\bibfield
  {journal} {\bibinfo  {journal} {Nature materials}\ }\textbf {\bibinfo
  {volume} {14}},\ \bibinfo {pages} {675} (\bibinfo {year} {2015})}\BibitemShut
  {NoStop}%
\bibitem [{\citenamefont {Raimondi}\ and\ \citenamefont {Schwab}(2005)}]{25}%
  \BibitemOpen
  \bibfield  {author} {\bibinfo {author} {\bibfnamefont {R.}~\bibnamefont
  {Raimondi}}\ and\ \bibinfo {author} {\bibfnamefont {P.}~\bibnamefont
  {Schwab}},\ }\href@noop {} {\bibfield  {journal} {\bibinfo  {journal}
  {Physical Review B}\ }\textbf {\bibinfo {volume} {71}},\ \bibinfo {pages}
  {033311} (\bibinfo {year} {2005})}\BibitemShut {NoStop}%
\bibitem [{\citenamefont {Milletar{\`\i}}\ and\ \citenamefont
  {Ferreira}(2016)}]{26}%
  \BibitemOpen
  \bibfield  {author} {\bibinfo {author} {\bibfnamefont {M.}~\bibnamefont
  {Milletar{\`\i}}}\ and\ \bibinfo {author} {\bibfnamefont {A.}~\bibnamefont
  {Ferreira}},\ }\href@noop {} {\bibfield  {journal} {\bibinfo  {journal}
  {Physical Review B}\ }\textbf {\bibinfo {volume} {94}},\ \bibinfo {pages}
  {134202} (\bibinfo {year} {2016})}\BibitemShut {NoStop}%
\bibitem [{\citenamefont {Ivchenko}\ and\ \citenamefont {Pikus}(1978)}]{27}%
  \BibitemOpen
  \bibfield  {author} {\bibinfo {author} {\bibfnamefont {E.}~\bibnamefont
  {Ivchenko}}\ and\ \bibinfo {author} {\bibfnamefont {G.}~\bibnamefont
  {Pikus}},\ }\href@noop {} {\bibfield  {journal} {\bibinfo  {journal} {Jetp
  Lett}\ }\textbf {\bibinfo {volume} {27}},\ \bibinfo {pages} {604} (\bibinfo
  {year} {1978})}\BibitemShut {NoStop}%
\bibitem [{\citenamefont {Vorob'ev}\ \emph {et~al.}(1979)\citenamefont
  {Vorob'ev}, \citenamefont {Ivchenko}, \citenamefont {Pikus}, \citenamefont
  {Farbshtein}, \citenamefont {Shalygin},\ and\ \citenamefont {Shturbin}}]{28}%
  \BibitemOpen
  \bibfield  {author} {\bibinfo {author} {\bibfnamefont {L.}~\bibnamefont
  {Vorob'ev}}, \bibinfo {author} {\bibfnamefont {E.}~\bibnamefont {Ivchenko}},
  \bibinfo {author} {\bibfnamefont {G.}~\bibnamefont {Pikus}}, \bibinfo
  {author} {\bibfnamefont {I.}~\bibnamefont {Farbshtein}}, \bibinfo {author}
  {\bibfnamefont {V.}~\bibnamefont {Shalygin}}, \ and\ \bibinfo {author}
  {\bibfnamefont {A.}~\bibnamefont {Shturbin}},\ }\href@noop {} {\bibfield
  {journal} {\bibinfo  {journal} {Soviet Journal of Experimental and
  Theoretical Physics Letters}\ }\textbf {\bibinfo {volume} {29}},\ \bibinfo
  {pages} {441} (\bibinfo {year} {1979})}\BibitemShut {NoStop}%
\bibitem [{\citenamefont {Ivchenko}\ \emph {et~al.}(1989)\citenamefont
  {Ivchenko}, \citenamefont {Lyanda-Geller},\ and\ \citenamefont {Pikus}}]{29}%
  \BibitemOpen
  \bibfield  {author} {\bibinfo {author} {\bibfnamefont {E.}~\bibnamefont
  {Ivchenko}}, \bibinfo {author} {\bibfnamefont {Y.~B.}\ \bibnamefont
  {Lyanda-Geller}}, \ and\ \bibinfo {author} {\bibfnamefont {G.}~\bibnamefont
  {Pikus}},\ }\href@noop {} {\bibfield  {journal} {\bibinfo  {journal} {JETP
  Lett}\ }\textbf {\bibinfo {volume} {50}},\ \bibinfo {pages} {175} (\bibinfo
  {year} {1989})}\BibitemShut {NoStop}%
\bibitem [{\citenamefont {Edelstein}(1990)}]{30}%
  \BibitemOpen
  \bibfield  {author} {\bibinfo {author} {\bibfnamefont {V.~M.}\ \bibnamefont
  {Edelstein}},\ }\href@noop {} {\bibfield  {journal} {\bibinfo  {journal}
  {Solid State Communications}\ }\textbf {\bibinfo {volume} {73}},\ \bibinfo
  {pages} {233} (\bibinfo {year} {1990})}\BibitemShut {NoStop}%
\bibitem [{\citenamefont {Aronov}\ and\ \citenamefont
  {Lyanda-Geller}(1989)}]{31}%
  \BibitemOpen
  \bibfield  {author} {\bibinfo {author} {\bibfnamefont {A.}~\bibnamefont
  {Aronov}}\ and\ \bibinfo {author} {\bibfnamefont {Y.~B.}\ \bibnamefont
  {Lyanda-Geller}},\ }\href@noop {} {\bibfield  {journal} {\bibinfo  {journal}
  {Soviet Journal of Experimental and Theoretical Physics Letters}\ }\textbf
  {\bibinfo {volume} {50}},\ \bibinfo {pages} {431} (\bibinfo {year}
  {1989})}\BibitemShut {NoStop}%
\bibitem [{\citenamefont {Rodriguez-Vega}\ \emph {et~al.}(2017)\citenamefont
  {Rodriguez-Vega}, \citenamefont {Schwiete}, \citenamefont {Sinova},\ and\
  \citenamefont {Rossi}}]{32}%
  \BibitemOpen
  \bibfield  {author} {\bibinfo {author} {\bibfnamefont {M.}~\bibnamefont
  {Rodriguez-Vega}}, \bibinfo {author} {\bibfnamefont {G.}~\bibnamefont
  {Schwiete}}, \bibinfo {author} {\bibfnamefont {J.}~\bibnamefont {Sinova}}, \
  and\ \bibinfo {author} {\bibfnamefont {E.}~\bibnamefont {Rossi}},\
  }\href@noop {} {\bibfield  {journal} {\bibinfo  {journal} {Physical Review
  B}\ }\textbf {\bibinfo {volume} {96}},\ \bibinfo {pages} {235419} (\bibinfo
  {year} {2017})}\BibitemShut {NoStop}%
\bibitem [{\citenamefont {Brataas}(2012)}]{33}%
  \BibitemOpen
  \bibfield  {author} {\bibinfo {author} {\bibfnamefont {A.}~\bibnamefont
  {Brataas}},\ }\href@noop {} {\bibfield  {journal} {\bibinfo  {journal} {Nat.
  Mater}\ }\textbf {\bibinfo {volume} {11}},\ \bibinfo {pages} {372} (\bibinfo
  {year} {2012})}\BibitemShut {NoStop}%
\bibitem [{\citenamefont {Bauer}\ \emph {et~al.}(2012)\citenamefont {Bauer},
  \citenamefont {Saitoh},\ and\ \citenamefont {Van~Wees}}]{34}%
  \BibitemOpen
  \bibfield  {author} {\bibinfo {author} {\bibfnamefont {G.~E.}\ \bibnamefont
  {Bauer}}, \bibinfo {author} {\bibfnamefont {E.}~\bibnamefont {Saitoh}}, \
  and\ \bibinfo {author} {\bibfnamefont {B.~J.}\ \bibnamefont {Van~Wees}},\
  }\href@noop {} {\bibfield  {journal} {\bibinfo  {journal} {Nature materials}\
  }\textbf {\bibinfo {volume} {11}},\ \bibinfo {pages} {391} (\bibinfo {year}
  {2012})}\BibitemShut {NoStop}%
\bibitem [{\citenamefont {Lesne}\ \emph {et~al.}(2016)\citenamefont {Lesne},
  \citenamefont {Fu}, \citenamefont {Oyarzun}, \citenamefont
  {Rojas-S{\'a}nchez}, \citenamefont {Vaz}, \citenamefont {Naganuma},
  \citenamefont {Sicoli}, \citenamefont {Attan{\'e}}, \citenamefont {Jamet},
  \citenamefont {Jacquet} \emph {et~al.}}]{36}%
  \BibitemOpen
  \bibfield  {author} {\bibinfo {author} {\bibfnamefont {E.}~\bibnamefont
  {Lesne}}, \bibinfo {author} {\bibfnamefont {Y.}~\bibnamefont {Fu}}, \bibinfo
  {author} {\bibfnamefont {S.}~\bibnamefont {Oyarzun}}, \bibinfo {author}
  {\bibfnamefont {J.}~\bibnamefont {Rojas-S{\'a}nchez}}, \bibinfo {author}
  {\bibfnamefont {D.}~\bibnamefont {Vaz}}, \bibinfo {author} {\bibfnamefont
  {H.}~\bibnamefont {Naganuma}}, \bibinfo {author} {\bibfnamefont
  {G.}~\bibnamefont {Sicoli}}, \bibinfo {author} {\bibfnamefont {J.-P.}\
  \bibnamefont {Attan{\'e}}}, \bibinfo {author} {\bibfnamefont
  {M.}~\bibnamefont {Jamet}}, \bibinfo {author} {\bibfnamefont
  {E.}~\bibnamefont {Jacquet}},  \emph {et~al.},\ }\href@noop {} {\bibfield
  {journal} {\bibinfo  {journal} {Nature materials}\ }\textbf {\bibinfo
  {volume} {15}},\ \bibinfo {pages} {1261} (\bibinfo {year}
  {2016})}\BibitemShut {NoStop}%
\bibitem [{\citenamefont {Shen}\ \emph {et~al.}(2014)\citenamefont {Shen},
  \citenamefont {Vignale},\ and\ \citenamefont {Raimondi}}]{37}%
  \BibitemOpen
  \bibfield  {author} {\bibinfo {author} {\bibfnamefont {K.}~\bibnamefont
  {Shen}}, \bibinfo {author} {\bibfnamefont {G.}~\bibnamefont {Vignale}}, \
  and\ \bibinfo {author} {\bibfnamefont {R.}~\bibnamefont {Raimondi}},\
  }\href@noop {} {\bibfield  {journal} {\bibinfo  {journal} {Physical review
  letters}\ }\textbf {\bibinfo {volume} {112}},\ \bibinfo {pages} {096601}
  (\bibinfo {year} {2014})}\BibitemShut {NoStop}%
\bibitem [{\citenamefont {Gorini}\ \emph {et~al.}(2017)\citenamefont {Gorini},
  \citenamefont {Sheikhabadi}, \citenamefont {Shen}, \citenamefont {Tokatly},
  \citenamefont {Vignale},\ and\ \citenamefont {Raimondi}}]{38}%
  \BibitemOpen
  \bibfield  {author} {\bibinfo {author} {\bibfnamefont {C.}~\bibnamefont
  {Gorini}}, \bibinfo {author} {\bibfnamefont {A.~M.}\ \bibnamefont
  {Sheikhabadi}}, \bibinfo {author} {\bibfnamefont {K.}~\bibnamefont {Shen}},
  \bibinfo {author} {\bibfnamefont {I.~V.}\ \bibnamefont {Tokatly}}, \bibinfo
  {author} {\bibfnamefont {G.}~\bibnamefont {Vignale}}, \ and\ \bibinfo
  {author} {\bibfnamefont {R.}~\bibnamefont {Raimondi}},\ }\href@noop {}
  {\bibfield  {journal} {\bibinfo  {journal} {Physical Review B}\ }\textbf
  {\bibinfo {volume} {95}},\ \bibinfo {pages} {205424} (\bibinfo {year}
  {2017})}\BibitemShut {NoStop}%
\bibitem [{\citenamefont {Huang}\ \emph {et~al.}(2017)\citenamefont {Huang},
  \citenamefont {Milletar{\`\i}},\ and\ \citenamefont {Cazalilla}}]{39}%
  \BibitemOpen
  \bibfield  {author} {\bibinfo {author} {\bibfnamefont {C.}~\bibnamefont
  {Huang}}, \bibinfo {author} {\bibfnamefont {M.}~\bibnamefont
  {Milletar{\`\i}}}, \ and\ \bibinfo {author} {\bibfnamefont {M.~A.}\
  \bibnamefont {Cazalilla}},\ }\href@noop {} {\bibfield  {journal} {\bibinfo
  {journal} {Physical Review B}\ }\textbf {\bibinfo {volume} {96}},\ \bibinfo
  {pages} {205305} (\bibinfo {year} {2017})}\BibitemShut {NoStop}%
\bibitem [{\citenamefont {Offidani}\ \emph {et~al.}(2017)\citenamefont
  {Offidani}, \citenamefont {Milletar{\`\i}}, \citenamefont {Raimondi},\ and\
  \citenamefont {Ferreira}}]{40}%
  \BibitemOpen
  \bibfield  {author} {\bibinfo {author} {\bibfnamefont {M.}~\bibnamefont
  {Offidani}}, \bibinfo {author} {\bibfnamefont {M.}~\bibnamefont
  {Milletar{\`\i}}}, \bibinfo {author} {\bibfnamefont {R.}~\bibnamefont
  {Raimondi}}, \ and\ \bibinfo {author} {\bibfnamefont {A.}~\bibnamefont
  {Ferreira}},\ }\href@noop {} {\bibfield  {journal} {\bibinfo  {journal}
  {Physical review letters}\ }\textbf {\bibinfo {volume} {119}},\ \bibinfo
  {pages} {196801} (\bibinfo {year} {2017})}\BibitemShut {NoStop}%
\bibitem [{\citenamefont {Mishchenko}\ \emph {et~al.}(2004)\citenamefont
  {Mishchenko}, \citenamefont {Shytov},\ and\ \citenamefont {Halperin}}]{41}%
  \BibitemOpen
  \bibfield  {author} {\bibinfo {author} {\bibfnamefont {E.~G.}\ \bibnamefont
  {Mishchenko}}, \bibinfo {author} {\bibfnamefont {A.~V.}\ \bibnamefont
  {Shytov}}, \ and\ \bibinfo {author} {\bibfnamefont {B.~I.}\ \bibnamefont
  {Halperin}},\ }\href@noop {} {\bibfield  {journal} {\bibinfo  {journal}
  {Physical review letters}\ }\textbf {\bibinfo {volume} {93}},\ \bibinfo
  {pages} {226602} (\bibinfo {year} {2004})}\BibitemShut {NoStop}%
\bibitem [{\citenamefont {Johansson}\ \emph {et~al.}(2016)\citenamefont
  {Johansson}, \citenamefont {Henk},\ and\ \citenamefont {Mertig}}]{42}%
  \BibitemOpen
  \bibfield  {author} {\bibinfo {author} {\bibfnamefont {A.}~\bibnamefont
  {Johansson}}, \bibinfo {author} {\bibfnamefont {J.}~\bibnamefont {Henk}}, \
  and\ \bibinfo {author} {\bibfnamefont {I.}~\bibnamefont {Mertig}},\
  }\href@noop {} {\bibfield  {journal} {\bibinfo  {journal} {Physical Review
  B}\ }\textbf {\bibinfo {volume} {93}},\ \bibinfo {pages} {195440} (\bibinfo
  {year} {2016})}\BibitemShut {NoStop}%
\bibitem [{\citenamefont {Akzyanov}(2019)}]{52}%
  \BibitemOpen
  \bibfield  {author} {\bibinfo {author} {\bibfnamefont {R.}~\bibnamefont
  {Akzyanov}},\ }\href@noop {} {\bibfield  {journal} {\bibinfo  {journal}
  {Physical Review B}\ }\textbf {\bibinfo {volume} {100}},\ \bibinfo {pages}
  {045403} (\bibinfo {year} {2019})}\BibitemShut {NoStop}%
\bibitem [{\citenamefont {Mecklenburg}\ and\ \citenamefont {Regan}(2011)}]{43}%
  \BibitemOpen
  \bibfield  {author} {\bibinfo {author} {\bibfnamefont {M.}~\bibnamefont
  {Mecklenburg}}\ and\ \bibinfo {author} {\bibfnamefont {B.}~\bibnamefont
  {Regan}},\ }\href@noop {} {\bibfield  {journal} {\bibinfo  {journal}
  {Physical Review Letters}\ }\textbf {\bibinfo {volume} {106}},\ \bibinfo
  {pages} {116803} (\bibinfo {year} {2011})}\BibitemShut {NoStop}%
\bibitem [{\citenamefont {Song}\ \emph {et~al.}(2015)\citenamefont {Song},
  \citenamefont {Paltoglou}, \citenamefont {Liu}, \citenamefont {Zhu},
  \citenamefont {Gallardo}, \citenamefont {Tang}, \citenamefont {Xu},
  \citenamefont {Ablowitz}, \citenamefont {Efremidis},\ and\ \citenamefont
  {Chen}}]{44}%
  \BibitemOpen
  \bibfield  {author} {\bibinfo {author} {\bibfnamefont {D.}~\bibnamefont
  {Song}}, \bibinfo {author} {\bibfnamefont {V.}~\bibnamefont {Paltoglou}},
  \bibinfo {author} {\bibfnamefont {S.}~\bibnamefont {Liu}}, \bibinfo {author}
  {\bibfnamefont {Y.}~\bibnamefont {Zhu}}, \bibinfo {author} {\bibfnamefont
  {D.}~\bibnamefont {Gallardo}}, \bibinfo {author} {\bibfnamefont
  {L.}~\bibnamefont {Tang}}, \bibinfo {author} {\bibfnamefont {J.}~\bibnamefont
  {Xu}}, \bibinfo {author} {\bibfnamefont {M.}~\bibnamefont {Ablowitz}},
  \bibinfo {author} {\bibfnamefont {N.~K.}\ \bibnamefont {Efremidis}}, \ and\
  \bibinfo {author} {\bibfnamefont {Z.}~\bibnamefont {Chen}},\ }\href@noop {}
  {\bibfield  {journal} {\bibinfo  {journal} {Nature communications}\ }\textbf
  {\bibinfo {volume} {6}},\ \bibinfo {pages} {6272} (\bibinfo {year}
  {2015})}\BibitemShut {NoStop}%
\bibitem [{\citenamefont {Geim}\ and\ \citenamefont {Novoselov}(2010)}]{45}%
  \BibitemOpen
  \bibfield  {author} {\bibinfo {author} {\bibfnamefont {A.~K.}\ \bibnamefont
  {Geim}}\ and\ \bibinfo {author} {\bibfnamefont {K.~S.}\ \bibnamefont
  {Novoselov}},\ }in\ \href@noop {} {\emph {\bibinfo {booktitle} {Nanoscience
  and Technology: A Collection of Reviews from Nature Journals}}}\ (\bibinfo
  {publisher} {World Scientific},\ \bibinfo {year} {2010})\ pp.\ \bibinfo
  {pages} {11--19}\BibitemShut {NoStop}%
\bibitem [{\citenamefont {Bostwick}\ \emph {et~al.}(2007)\citenamefont
  {Bostwick}, \citenamefont {Ohta}, \citenamefont {Seyller}, \citenamefont
  {Horn},\ and\ \citenamefont {Rotenberg}}]{46}%
  \BibitemOpen
  \bibfield  {author} {\bibinfo {author} {\bibfnamefont {A.}~\bibnamefont
  {Bostwick}}, \bibinfo {author} {\bibfnamefont {T.}~\bibnamefont {Ohta}},
  \bibinfo {author} {\bibfnamefont {T.}~\bibnamefont {Seyller}}, \bibinfo
  {author} {\bibfnamefont {K.}~\bibnamefont {Horn}}, \ and\ \bibinfo {author}
  {\bibfnamefont {E.}~\bibnamefont {Rotenberg}},\ }\href@noop {} {\bibfield
  {journal} {\bibinfo  {journal} {Nature physics}\ }\textbf {\bibinfo {volume}
  {3}},\ \bibinfo {pages} {36} (\bibinfo {year} {2007})}\BibitemShut {NoStop}%
\bibitem [{\citenamefont {Zhou}\ \emph {et~al.}(2007)\citenamefont {Zhou},
  \citenamefont {Gweon}, \citenamefont {Fedorov}, \citenamefont {First},
  \citenamefont {De~Heer}, \citenamefont {Lee}, \citenamefont {Guinea},
  \citenamefont {Neto},\ and\ \citenamefont {Lanzara}}]{47}%
  \BibitemOpen
  \bibfield  {author} {\bibinfo {author} {\bibfnamefont {S.~Y.}\ \bibnamefont
  {Zhou}}, \bibinfo {author} {\bibfnamefont {G.-H.}\ \bibnamefont {Gweon}},
  \bibinfo {author} {\bibfnamefont {A.}~\bibnamefont {Fedorov}}, \bibinfo
  {author} {\bibfnamefont {d.}~\bibnamefont {First}, \bibfnamefont {PN}},
  \bibinfo {author} {\bibfnamefont {W.}~\bibnamefont {De~Heer}}, \bibinfo
  {author} {\bibfnamefont {D.-H.}\ \bibnamefont {Lee}}, \bibinfo {author}
  {\bibfnamefont {F.}~\bibnamefont {Guinea}}, \bibinfo {author} {\bibfnamefont
  {A.~C.}\ \bibnamefont {Neto}}, \ and\ \bibinfo {author} {\bibfnamefont
  {A.}~\bibnamefont {Lanzara}},\ }\href@noop {} {\bibfield  {journal} {\bibinfo
   {journal} {Nature materials}\ }\textbf {\bibinfo {volume} {6}},\ \bibinfo
  {pages} {770} (\bibinfo {year} {2007})}\BibitemShut {NoStop}%
\bibitem [{\citenamefont {Nair}\ \emph {et~al.}(2008)\citenamefont {Nair},
  \citenamefont {Blake}, \citenamefont {Grigorenko}, \citenamefont {Novoselov},
  \citenamefont {Booth}, \citenamefont {Stauber}, \citenamefont {Peres},\ and\
  \citenamefont {Geim}}]{48}%
  \BibitemOpen
  \bibfield  {author} {\bibinfo {author} {\bibfnamefont {R.~R.}\ \bibnamefont
  {Nair}}, \bibinfo {author} {\bibfnamefont {P.}~\bibnamefont {Blake}},
  \bibinfo {author} {\bibfnamefont {A.~N.}\ \bibnamefont {Grigorenko}},
  \bibinfo {author} {\bibfnamefont {K.~S.}\ \bibnamefont {Novoselov}}, \bibinfo
  {author} {\bibfnamefont {T.~J.}\ \bibnamefont {Booth}}, \bibinfo {author}
  {\bibfnamefont {T.}~\bibnamefont {Stauber}}, \bibinfo {author} {\bibfnamefont
  {N.~M.}\ \bibnamefont {Peres}}, \ and\ \bibinfo {author} {\bibfnamefont
  {A.~K.}\ \bibnamefont {Geim}},\ }\href@noop {} {\bibfield  {journal}
  {\bibinfo  {journal} {Science}\ }\textbf {\bibinfo {volume} {320}},\ \bibinfo
  {pages} {1308} (\bibinfo {year} {2008})}\BibitemShut {NoStop}%
\bibitem [{\citenamefont {Pesin}\ and\ \citenamefont {MacDonald}(2012)}]{49}%
  \BibitemOpen
  \bibfield  {author} {\bibinfo {author} {\bibfnamefont {D.}~\bibnamefont
  {Pesin}}\ and\ \bibinfo {author} {\bibfnamefont {A.~H.}\ \bibnamefont
  {MacDonald}},\ }\href@noop {} {\bibfield  {journal} {\bibinfo  {journal}
  {Nature materials}\ }\textbf {\bibinfo {volume} {11}},\ \bibinfo {pages}
  {409} (\bibinfo {year} {2012})}\BibitemShut {NoStop}%
\bibitem [{\citenamefont {Chen}\ \emph {et~al.}(2014)\citenamefont {Chen},
  \citenamefont {Li},\ and\ \citenamefont {Wang}}]{50}%
  \BibitemOpen
  \bibfield  {author} {\bibinfo {author} {\bibfnamefont {J.}~\bibnamefont
  {Chen}}, \bibinfo {author} {\bibfnamefont {Z.}~\bibnamefont {Li}}, \ and\
  \bibinfo {author} {\bibfnamefont {W.}~\bibnamefont {Wang}},\ }\href@noop {}
  {\bibfield  {journal} {\bibinfo  {journal} {Journal of Applied Physics}\
  }\textbf {\bibinfo {volume} {115}},\ \bibinfo {pages} {053701} (\bibinfo
  {year} {2014})}\BibitemShut {NoStop}%
\bibitem [{\citenamefont {Johansson}\ \emph {et~al.}(2018)\citenamefont
  {Johansson}, \citenamefont {Henk},\ and\ \citenamefont {Mertig}}]{35}%
  \BibitemOpen
  \bibfield  {author} {\bibinfo {author} {\bibfnamefont {A.}~\bibnamefont
  {Johansson}}, \bibinfo {author} {\bibfnamefont {J.}~\bibnamefont {Henk}}, \
  and\ \bibinfo {author} {\bibfnamefont {I.}~\bibnamefont {Mertig}},\
  }\href@noop {} {\bibfield  {journal} {\bibinfo  {journal} {Physical Review
  B}\ }\textbf {\bibinfo {volume} {97}},\ \bibinfo {pages} {085417} (\bibinfo
  {year} {2018})}\BibitemShut {NoStop}%
\bibitem [{\citenamefont {Hajzadeh}\ \emph {et~al.}(2019)\citenamefont
  {Hajzadeh}, \citenamefont {Rahmati}, \citenamefont {Jafari},\ and\
  \citenamefont {Mohseni}}]{53}%
  \BibitemOpen
  \bibfield  {author} {\bibinfo {author} {\bibfnamefont {I.}~\bibnamefont
  {Hajzadeh}}, \bibinfo {author} {\bibfnamefont {B.}~\bibnamefont {Rahmati}},
  \bibinfo {author} {\bibfnamefont {G.}~\bibnamefont {Jafari}}, \ and\ \bibinfo
  {author} {\bibfnamefont {S.}~\bibnamefont {Mohseni}},\ }\href@noop {}
  {\bibfield  {journal} {\bibinfo  {journal} {Physical Review B}\ }\textbf
  {\bibinfo {volume} {99}},\ \bibinfo {pages} {094414} (\bibinfo {year}
  {2019})}\BibitemShut {NoStop}%
\bibitem [{\citenamefont {Luan}\ \emph {et~al.}(2019)\citenamefont {Luan},
  \citenamefont {Zhou}, \citenamefont {Wang}, \citenamefont {Zhang},
  \citenamefont {Du}, \citenamefont {Xiao}, \citenamefont {Liu},\ and\
  \citenamefont {Wu}}]{54}%
  \BibitemOpen
  \bibfield  {author} {\bibinfo {author} {\bibfnamefont {Z.}~\bibnamefont
  {Luan}}, \bibinfo {author} {\bibfnamefont {L.}~\bibnamefont {Zhou}}, \bibinfo
  {author} {\bibfnamefont {P.}~\bibnamefont {Wang}}, \bibinfo {author}
  {\bibfnamefont {S.}~\bibnamefont {Zhang}}, \bibinfo {author} {\bibfnamefont
  {J.}~\bibnamefont {Du}}, \bibinfo {author} {\bibfnamefont {J.}~\bibnamefont
  {Xiao}}, \bibinfo {author} {\bibfnamefont {R.}~\bibnamefont {Liu}}, \ and\
  \bibinfo {author} {\bibfnamefont {D.}~\bibnamefont {Wu}},\ }\href@noop {}
  {\bibfield  {journal} {\bibinfo  {journal} {Physical Review B}\ }\textbf
  {\bibinfo {volume} {99}},\ \bibinfo {pages} {174406} (\bibinfo {year}
  {2019})}\BibitemShut {NoStop}%
\bibitem [{\citenamefont {Skowro{\'n}ski}\ and\ \citenamefont {et.
  al.}(2019)}]{55}%
  \BibitemOpen
  \bibfield  {author} {\bibinfo {author} {\bibnamefont {Skowro{\'n}ski}}\ and\
  \bibinfo {author} {\bibnamefont {et. al.}},\ }\href@noop {} {\bibfield
  {journal} {\bibinfo  {journal} {Physical Review Applied}\ }\textbf {\bibinfo
  {volume} {11}},\ \bibinfo {pages} {024039} (\bibinfo {year}
  {2019})}\BibitemShut {NoStop}%
\bibitem [{\citenamefont {Zhou}\ \emph {et~al.}(2019)\citenamefont {Zhou},
  \citenamefont {Qiao}, \citenamefont {Bournel},\ and\ \citenamefont
  {Zhao}}]{56}%
  \BibitemOpen
  \bibfield  {author} {\bibinfo {author} {\bibfnamefont {J.}~\bibnamefont
  {Zhou}}, \bibinfo {author} {\bibfnamefont {J.}~\bibnamefont {Qiao}}, \bibinfo
  {author} {\bibfnamefont {A.}~\bibnamefont {Bournel}}, \ and\ \bibinfo
  {author} {\bibfnamefont {W.}~\bibnamefont {Zhao}},\ }\href@noop {} {\bibfield
   {journal} {\bibinfo  {journal} {Physical Review B}\ }\textbf {\bibinfo
  {volume} {99}},\ \bibinfo {pages} {060408} (\bibinfo {year}
  {2019})}\BibitemShut {NoStop}%
\bibitem [{\citenamefont {Cramer}\ \emph {et~al.}(2019)\citenamefont {Cramer},
  \citenamefont {Ross}, \citenamefont {Jaiswal}, \citenamefont {Baldrati},
  \citenamefont {Lebrun},\ and\ \citenamefont {Kl{\"a}ui}}]{57}%
  \BibitemOpen
  \bibfield  {author} {\bibinfo {author} {\bibfnamefont {J.}~\bibnamefont
  {Cramer}}, \bibinfo {author} {\bibfnamefont {A.}~\bibnamefont {Ross}},
  \bibinfo {author} {\bibfnamefont {S.}~\bibnamefont {Jaiswal}}, \bibinfo
  {author} {\bibfnamefont {L.}~\bibnamefont {Baldrati}}, \bibinfo {author}
  {\bibfnamefont {R.}~\bibnamefont {Lebrun}}, \ and\ \bibinfo {author}
  {\bibfnamefont {M.}~\bibnamefont {Kl{\"a}ui}},\ }\href@noop {} {\bibfield
  {journal} {\bibinfo  {journal} {Physical Review B}\ }\textbf {\bibinfo
  {volume} {99}},\ \bibinfo {pages} {104414} (\bibinfo {year}
  {2019})}\BibitemShut {NoStop}%
\bibitem [{\citenamefont {Peters}\ and\ \citenamefont {Yanase}(2018)}]{58}%
  \BibitemOpen
  \bibfield  {author} {\bibinfo {author} {\bibfnamefont {R.}~\bibnamefont
  {Peters}}\ and\ \bibinfo {author} {\bibfnamefont {Y.}~\bibnamefont
  {Yanase}},\ }\href@noop {} {\bibfield  {journal} {\bibinfo  {journal}
  {Physical Review B}\ }\textbf {\bibinfo {volume} {97}},\ \bibinfo {pages}
  {115128} (\bibinfo {year} {2018})}\BibitemShut {NoStop}%
\bibitem [{\citenamefont {Sheikhabadi}\ \emph {et~al.}(2018)\citenamefont
  {Sheikhabadi}, \citenamefont {Miatka}, \citenamefont {Sherman},\ and\
  \citenamefont {Raimondi}}]{59}%
  \BibitemOpen
  \bibfield  {author} {\bibinfo {author} {\bibfnamefont {A.~M.}\ \bibnamefont
  {Sheikhabadi}}, \bibinfo {author} {\bibfnamefont {I.}~\bibnamefont {Miatka}},
  \bibinfo {author} {\bibfnamefont {E.~Y.}\ \bibnamefont {Sherman}}, \ and\
  \bibinfo {author} {\bibfnamefont {R.}~\bibnamefont {Raimondi}},\ }\href@noop
  {} {\bibfield  {journal} {\bibinfo  {journal} {Physical Review B}\ }\textbf
  {\bibinfo {volume} {97}},\ \bibinfo {pages} {235412} (\bibinfo {year}
  {2018})}\BibitemShut {NoStop}%
\bibitem [{\citenamefont {Xu}\ \emph {et~al.}(2018)\citenamefont {Xu},
  \citenamefont {Puebla}, \citenamefont {Auvray}, \citenamefont {Rana},
  \citenamefont {Kondou},\ and\ \citenamefont {Otani}}]{60}%
  \BibitemOpen
  \bibfield  {author} {\bibinfo {author} {\bibfnamefont {M.}~\bibnamefont
  {Xu}}, \bibinfo {author} {\bibfnamefont {J.}~\bibnamefont {Puebla}}, \bibinfo
  {author} {\bibfnamefont {F.}~\bibnamefont {Auvray}}, \bibinfo {author}
  {\bibfnamefont {B.}~\bibnamefont {Rana}}, \bibinfo {author} {\bibfnamefont
  {K.}~\bibnamefont {Kondou}}, \ and\ \bibinfo {author} {\bibfnamefont
  {Y.}~\bibnamefont {Otani}},\ }\href@noop {} {\bibfield  {journal} {\bibinfo
  {journal} {Physical Review B}\ }\textbf {\bibinfo {volume} {97}},\ \bibinfo
  {pages} {180301} (\bibinfo {year} {2018})}\BibitemShut {NoStop}%
\bibitem [{\citenamefont {Zhou}\ \emph {et~al.}(2018)\citenamefont {Zhou},
  \citenamefont {Liu}, \citenamefont {Wang}, \citenamefont {Ma}, \citenamefont
  {Jia}, \citenamefont {Wu}, \citenamefont {Zhou}, \citenamefont {Zhang},
  \citenamefont {Liu}, \citenamefont {Wu} \emph {et~al.}}]{61}%
  \BibitemOpen
  \bibfield  {author} {\bibinfo {author} {\bibfnamefont {C.}~\bibnamefont
  {Zhou}}, \bibinfo {author} {\bibfnamefont {Y.}~\bibnamefont {Liu}}, \bibinfo
  {author} {\bibfnamefont {Z.}~\bibnamefont {Wang}}, \bibinfo {author}
  {\bibfnamefont {S.}~\bibnamefont {Ma}}, \bibinfo {author} {\bibfnamefont
  {M.}~\bibnamefont {Jia}}, \bibinfo {author} {\bibfnamefont {R.}~\bibnamefont
  {Wu}}, \bibinfo {author} {\bibfnamefont {L.}~\bibnamefont {Zhou}}, \bibinfo
  {author} {\bibfnamefont {W.}~\bibnamefont {Zhang}}, \bibinfo {author}
  {\bibfnamefont {M.}~\bibnamefont {Liu}}, \bibinfo {author} {\bibfnamefont
  {Y.}~\bibnamefont {Wu}},  \emph {et~al.},\ }\href@noop {} {\bibfield
  {journal} {\bibinfo  {journal} {Physical review letters}\ }\textbf {\bibinfo
  {volume} {121}},\ \bibinfo {pages} {086801} (\bibinfo {year}
  {2018})}\BibitemShut {NoStop}%
\bibitem [{\citenamefont {Massarelli}\ \emph {et~al.}(2019)\citenamefont
  {Massarelli}, \citenamefont {Wu},\ and\ \citenamefont {Paramekanti}}]{62}%
  \BibitemOpen
  \bibfield  {author} {\bibinfo {author} {\bibfnamefont {G.}~\bibnamefont
  {Massarelli}}, \bibinfo {author} {\bibfnamefont {B.}~\bibnamefont {Wu}}, \
  and\ \bibinfo {author} {\bibfnamefont {A.}~\bibnamefont {Paramekanti}},\
  }\href@noop {} {\bibfield  {journal} {\bibinfo  {journal} {arXiv preprint
  arXiv:1904.04280}\ } (\bibinfo {year} {2019})}\BibitemShut {NoStop}%
\bibitem [{\citenamefont {Ezawa}(2012{\natexlab{a}})}]{63}%
  \BibitemOpen
  \bibfield  {author} {\bibinfo {author} {\bibfnamefont {M.}~\bibnamefont
  {Ezawa}},\ }\href@noop {} {\bibfield  {journal} {\bibinfo  {journal} {New
  Journal of Physics}\ }\textbf {\bibinfo {volume} {14}},\ \bibinfo {pages}
  {033003} (\bibinfo {year} {2012}{\natexlab{a}})}\BibitemShut {NoStop}%
\bibitem [{\citenamefont {Ezawa}(2012{\natexlab{b}})}]{64}%
  \BibitemOpen
  \bibfield  {author} {\bibinfo {author} {\bibfnamefont {M.}~\bibnamefont
  {Ezawa}},\ }\href@noop {} {\bibfield  {journal} {\bibinfo  {journal} {Journal
  of the Physical Society of Japan}\ }\textbf {\bibinfo {volume} {81}},\
  \bibinfo {pages} {064705} (\bibinfo {year} {2012}{\natexlab{b}})}\BibitemShut
  {NoStop}%
\bibitem [{\citenamefont {Cr{\'e}pieux}\ and\ \citenamefont
  {Bruno}(2001)}]{65}%
  \BibitemOpen
  \bibfield  {author} {\bibinfo {author} {\bibfnamefont {A.}~\bibnamefont
  {Cr{\'e}pieux}}\ and\ \bibinfo {author} {\bibfnamefont {P.}~\bibnamefont
  {Bruno}},\ }\href@noop {} {\bibfield  {journal} {\bibinfo  {journal}
  {Physical Review B}\ }\textbf {\bibinfo {volume} {64}},\ \bibinfo {pages}
  {014416} (\bibinfo {year} {2001})}\BibitemShut {NoStop}%
\end{thebibliography}%
\end{document}